# Inverted analemmatic sundial of the Bronze Age

**Larisa N. Vodolazhskaya**[1]

**Abstract**

This article describes the study of a Bronze Age limestone slab with cup marks, discovered during the archaeological excavations of kurgan 1 of the kurgan grave field Prolom II, located in the Belogorsk region in Crimea. According to the results of the study, it is concluded that the Belogorsk slab is a sundial of about the XV - XII centuries BC and belongs to the Srubnaya culture. By type of sundial, it is closest to the analemmatic sundial. However, the principle of the hourly markings of the Belogorsk slab is so unique that it is proposed to separate this type of sundial into a new type - inverted analemmatic sundial. This type is characterized by the fact that, unlike typical analemmatic sundials, the gnomon remains motionless throughout the year, and in accordance with the analemma, the "dial" "moves" - an ellipse of hour markers from cup marks, i.e. gnomon and hour markers (cup marks) change places in terms of mobility. The movement of hour markers on the Belogorsk slab is not literal, but is imitated by several rows of cup marks, which are fragments of hour marker ellipses for different months of the year. The idea behind this type of sundial is so revolutionary that we can talk about the discovery of a completely new type of sundial, the analogue of which has not yet been discovered.

In 2017, the expedition of the museum-reserve "Scythian Neapolis" in the process of archaeological excavations of kurgan 1 of the kurgan grave field Prolom II (1,7 km east of the Prolom village), located in the Belogorsk district of Crimea. A break, among the blockage of the stone shell of the kurgan, a stone with hollowed cup marks was found (Fig. 1)[2]. It was located among the blockage of a stone carapace, which covered the top of a stone burial structure of the late 4th century BC, let into the kurgan grave field of the Bronze Age (earth pit with a burial, later completely robbed; most likely, ocher was present in the burial). The stone was dated to the Bronze Age.

The stone is a limestone slab of subrectangular shape. On one of the sides, rounded cup marks are knocked out, most of which have a diameter of about 1-1,5 cm and a depth of about 0,5 cm.

The main feature of the slab is small cup marks densely located along the edge of the slab and partially filling its inner area. On one side of the slab along its edge, cup marks are applied, forming a curved line resembling the arc of an ellipse.

To date, three slabs of fine-grained sandstone with cup marks arranged in a circle, which belong to the Late Bronze Age, as well as a slab of coarse-grained sandstone with a semicircular groove, which are attributed to the Early Bronze Age, have been found in the Northern Black Sea region.

---

[1] E-mails: larisavodol@aaatec.org, larisavodol@gmail.com

[2] Museum inventory number KM 7265/5 (Historical and archaeological museum-reserve "Scythian Neapolis").

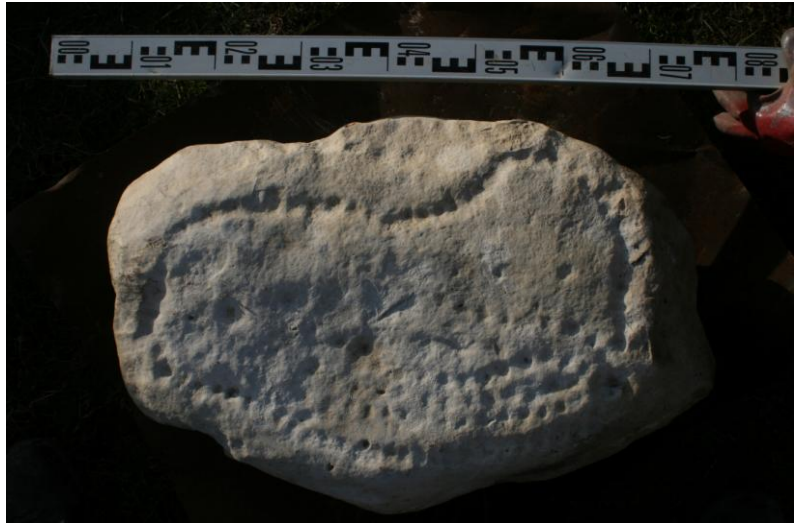

**Figure 1.** Kurgan grave field Prolom II, kurgan 1, slab with cup marks (photo by V. Nuzhdenko, 2017).

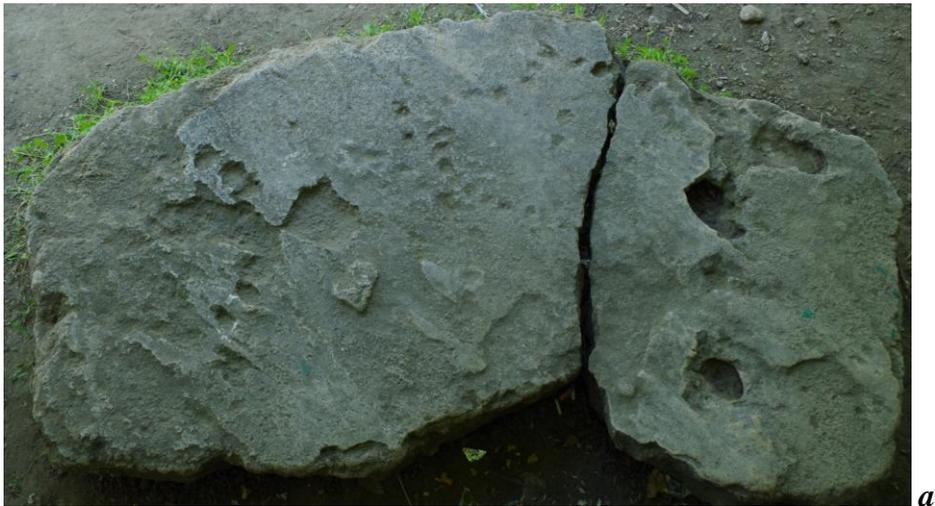

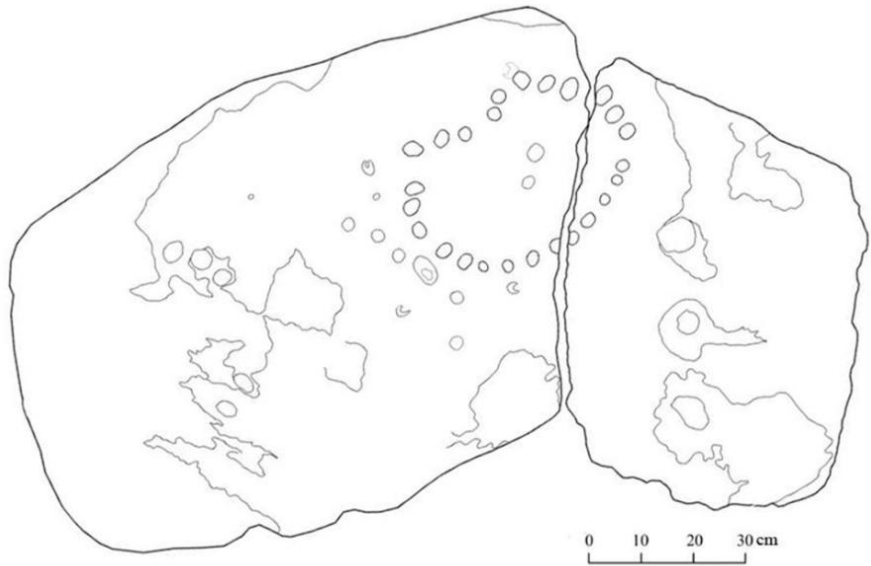

**Figure 2.** Kurgan field Tavriya-1, kurgan 1, burial 2, slab with cup marks: ***a*** – photograph of the surface of the slab with cup marks (Vodolazhskaya, Larenok, Nevsky, 2014, Fig. 3); ***b*** – drawing the surface of a slab with cup marks (Vodolazhskaya, Larenok, Nevsky, 2014, Fig. 4b).

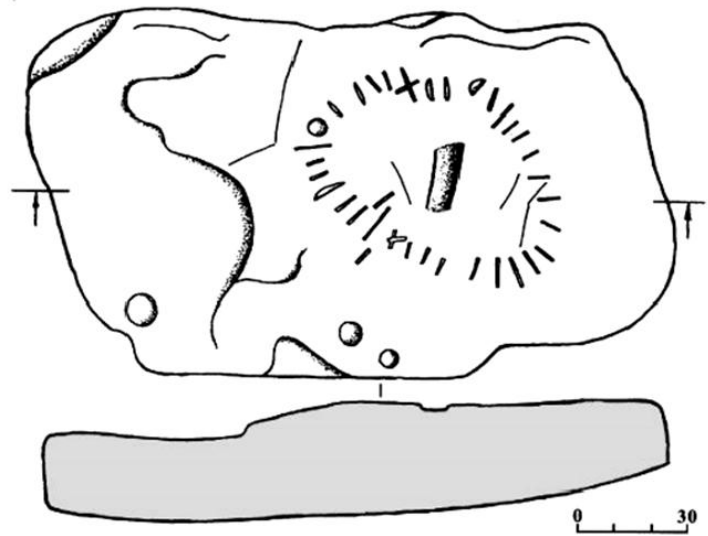

**Figure 3.** Kurgan group Rusin Yar, kurgan 1, burial 1, drawing of a slab with cup marks (Polidovych, Usachuk, 2013, Fig. 2.1).

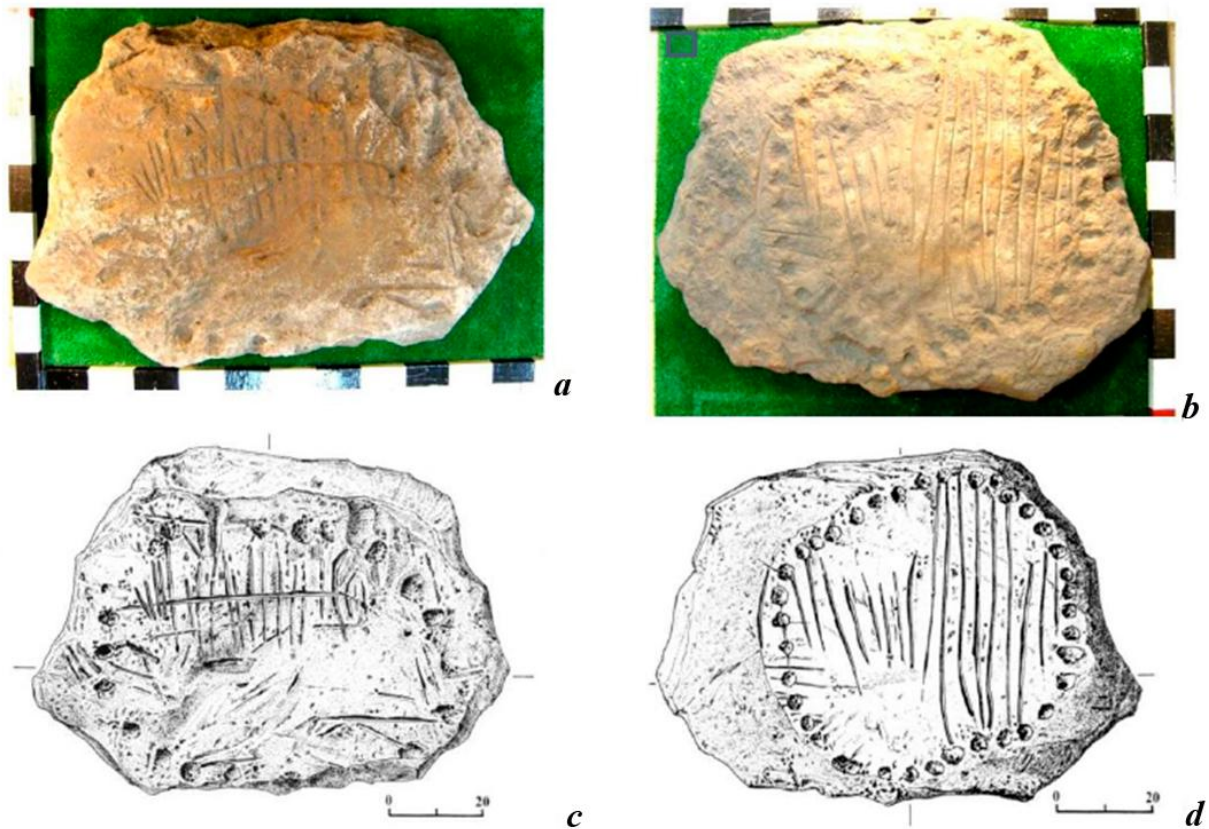

**Figure 4.** Kurgan group Popov Yar-2, kurgan 3, burial 7, slab with cup marks: ***a*** – photo of side A (Polidovich et al., 2012, Fig. 214), ***b*** – photo of side B (Polidovich et al., 2012, Fig. 218), ***c*** – drawing of side A (Polidovich et al., 2012, Fig. 212), ***d*** – drawing of side B (Polidovich et al., 2012, Fig. 212).

A slab with cup marks from the Tavriya-1 grave field in the Rostov region is stored on the territory of the Tanais Archaeological Museum-Reserve (Rostov region, Russia) (Larenok, 1998, p. 62) (Fig. 2). It dates from the XV-XIII centuries BC and attributed to the Srubnaya culture. Two slabs with cup marks - from the kurgan groups Rusin Yar (heavily damaged later) (Fig. 3) and Popov Yar-2 of the Donetsk region (Polidovych, Usachuk, 2013, p. 53-67; Polidovich,

Usachuk, 2015, p. 444, 455; Polidovich, et al., 2013, p. 36-135) (Fig. 4) are in the Konstantinovsky City Museum and in the Donetsk Regional Museum of Local Lore respectively. They date from the XIII-XII centuries BC and belong to the Srubnaya culture.

A slab with a groove shaped like a semicircle was found near kurgan 1 of the Varvarinsky I grave field in the Rostov Region (Faifert, 2015, p. 27-28) and is stored in the Azov Historical, Archaeological and Paleontological Museum-Reserve (Rostov Region, Russia) (Fig. 5). The slab is dated to the end of the IV millennium BC and belongs to the Yamnaya culture.

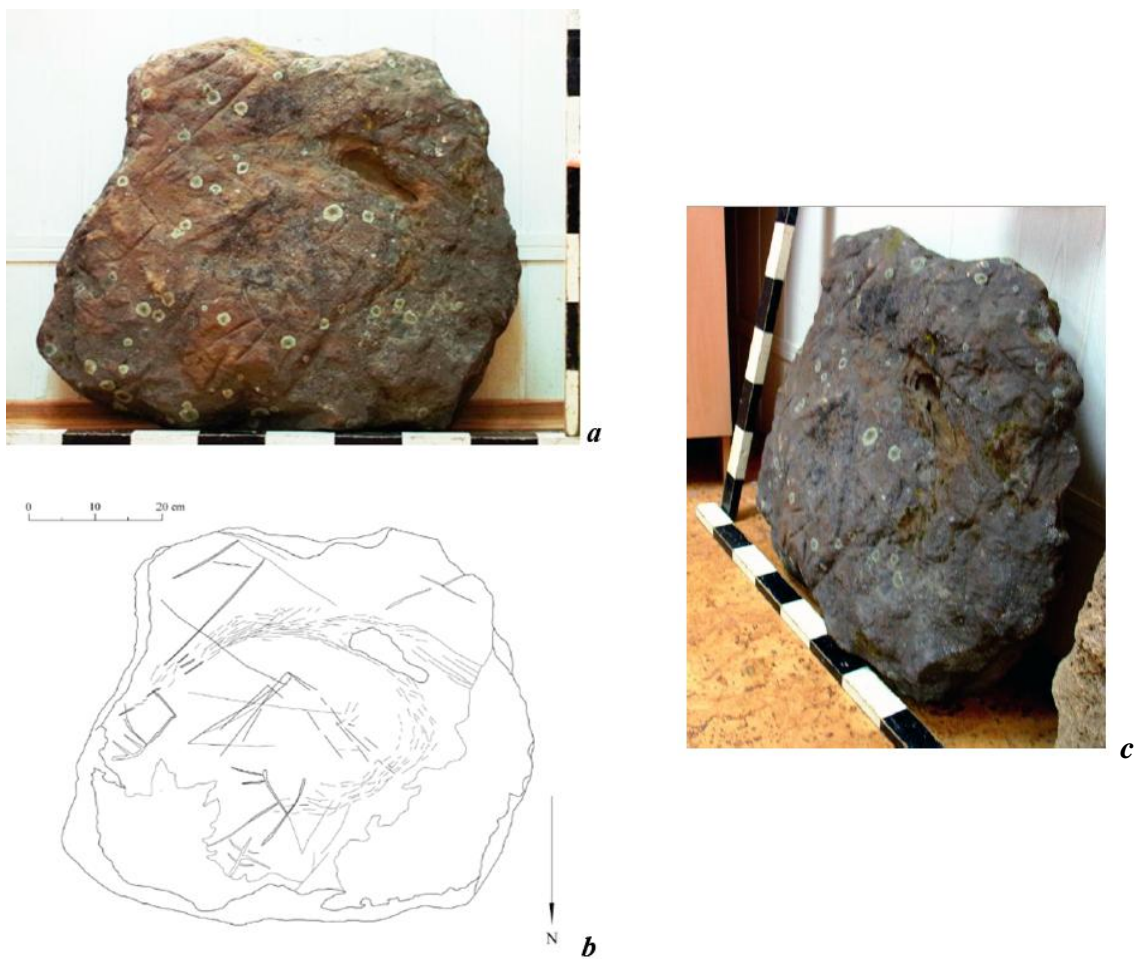

**Figure 5.** Kurgan field Varvarinsky I, kurgan 1 (neighborhood), a slab with a semicircle-shaped groove: ***a*** – photograph of the slab (Vodolazhskaya, Larenok, Nevsky, 2016, Fig. 7a); ***b*** – drawing of the slab (Vodolazhskaya, Larenok, Nevsky, 2016, Fig. 7b); ***c*** – photograph of the slab (side view) (Vodolazhskaya, Larenok, Nevsky, 2016, Fig. 11).

Studies of the above slabs have shown that the cup marks on these slabs represent the hour markers of the analemmatic sundial, and in the case of the Varvarinsky slab, an ellipse of hour markers (Vodolazhskaya, 2013, p. 68-88; Vodolazhskaya, Larenok, Nevsky, 2014, p. 31-53; Vodolazhskaya, Larenok, Nevsky, 2016, p. 96-116; Vodolazhskaya, Larenok, Nevsky, 2016, p. 150-168).

Another slab with cup marks located in a semicircle and associated with the marking of analemmatic sundial was discovered in the Krasnodar Krai near a heavily plowed kurgan near the village of Pyatikhatki, Anapa region (Novichikhin, 1995, p. 25-27; Vodolazhskaya, Novichikhin, Nevsky, 2021, p. 73-86). Now it is stored in the Anapa Archaeological Museum (Fig. 6). It was attributed to the Dolmen culture and dated to the XXV-XV centuries BC.

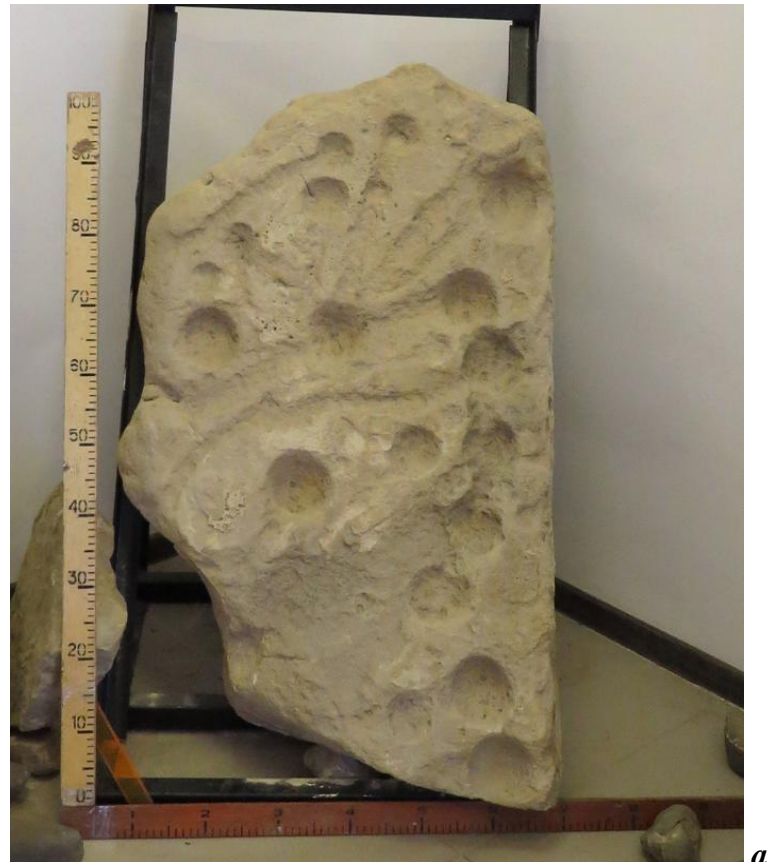

*a*

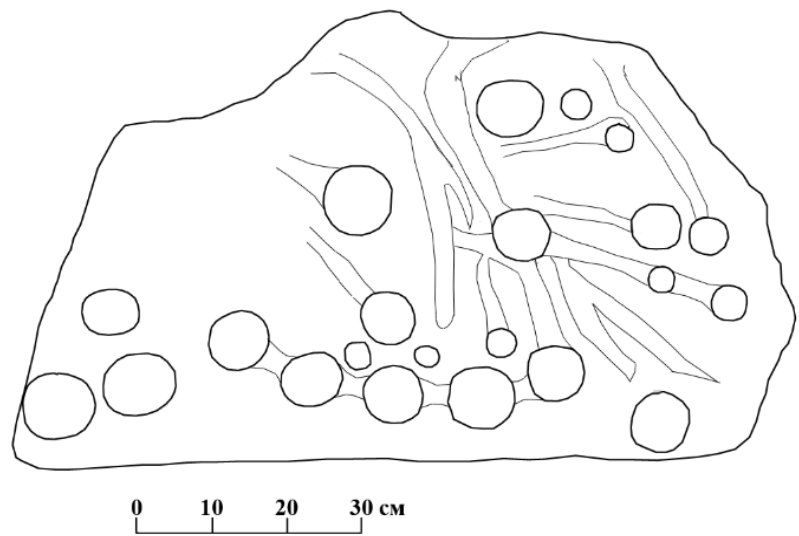

*b*

**Figure 6.** Settlement Pyatikhatki, slab with cup marks: ***a*** – photo of the slab in the modern exposition of the Anapa Archaeological Museum (Vodolazhskaya, Novichikhin, Nevsky, 2021, Fig. 1b); ***b*** – schematic drawing of the surface of the slab (Vodolazhskaya, Novichikhin, Nevsky, 2021, Fig. 3b).

Because on the Belogorsk slab, the cup marks on one side are arranged in an arc similar to an ellipse, it was suggested that the slab could be a slab with hour markers of an analemmatic sundial. In this case, taking into account the nature of the cup marks and their size, the slab can be attributed to the Srubnaya culture by analogy with the slabs found in the kurgan field Tavriya-1 (Fig. 2) and in the kurgan group Popov Yar-2 (Fig. 4).

For a more detailed analysis of the features of the location of cup marks on the surface of the Belogorsk slab, a standard photo correction was carried out: distortions that occurred at the edges of the photograph were corrected, contrast and brightness were increased, and a schematic drawing of the most clearly visible cup marks on the surface of the slab was made (Fig. 7).

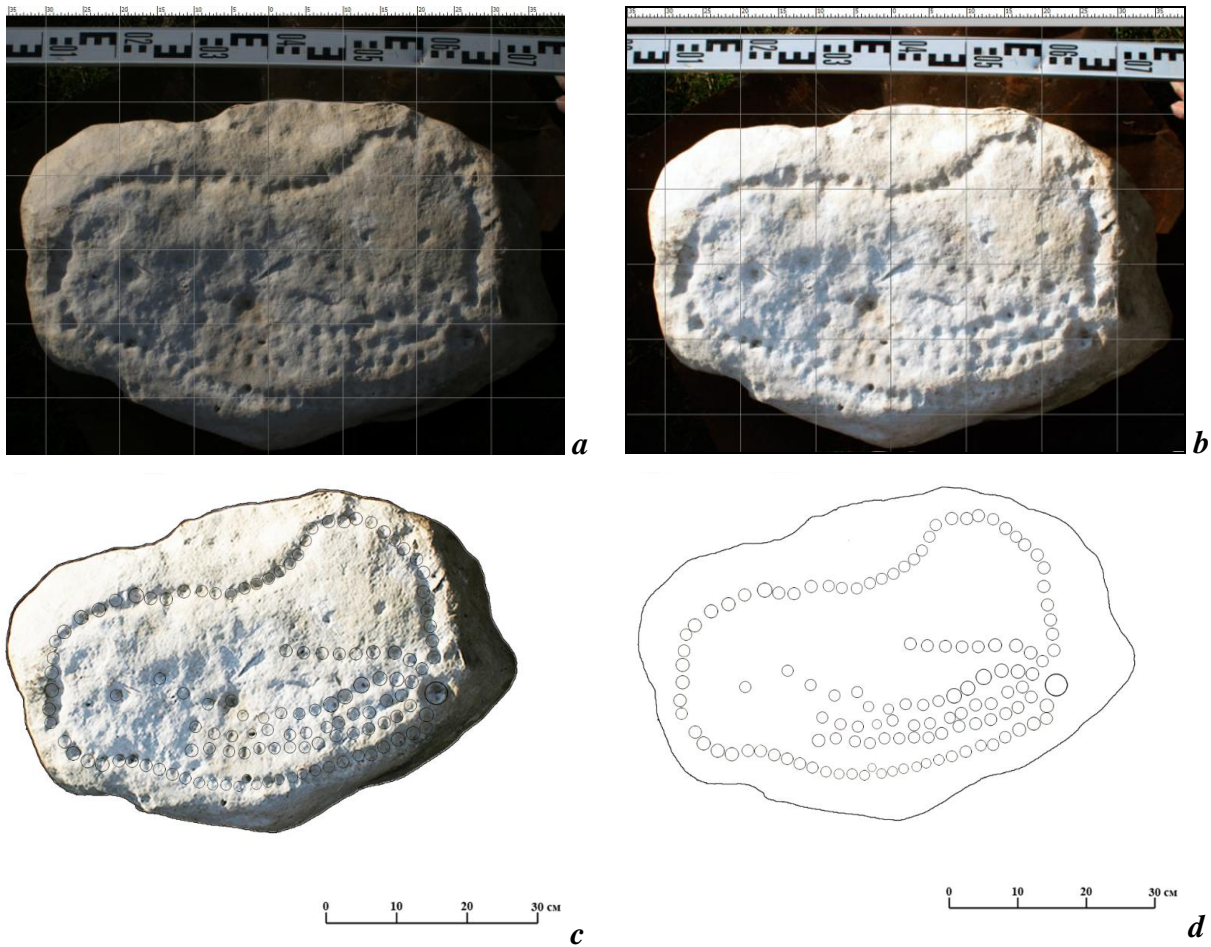

**Figure 7.** Kurgan grave field Prolom II, kurgan 1, slab with cup marks: *a* – photograph; *b* – photograph after correction; *c* – corrected photograph with a schematic drawing of the cup marks; *d* – schematic drawing of cup marks on the surface of the slab.

Only a part of the cup marks on the surface of the Belogorsk slab is made in the same style and can be unambiguously considered as a complex of cup marks. These cup marks have the clearest shape and are close in size. It was this complex of cup marks that was considered from the point of view of the possible marking of analemmatic sundials. Another, much smaller part of the cup marks, has a smaller depth and fuzzy shape. We considered these cup marks as random and did not analyze them. The differences between these types of cup marks are clearly visible in Figure 8.

In the case of an analemmatic sundial, the cup marks along the ellipse should be in the northern part of the slab. The slab should be placed horizontally, and the gnomon vertically. The gnomon must be installed and moved along the surface of the slab according to the principle of a chess piece, or suspended above the slab in the form of a plumb line.

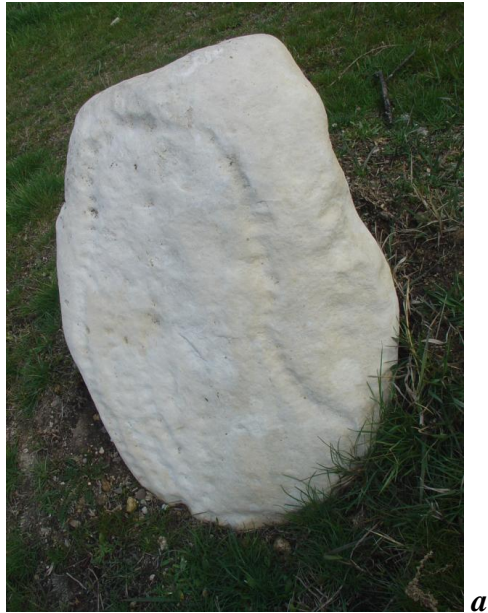

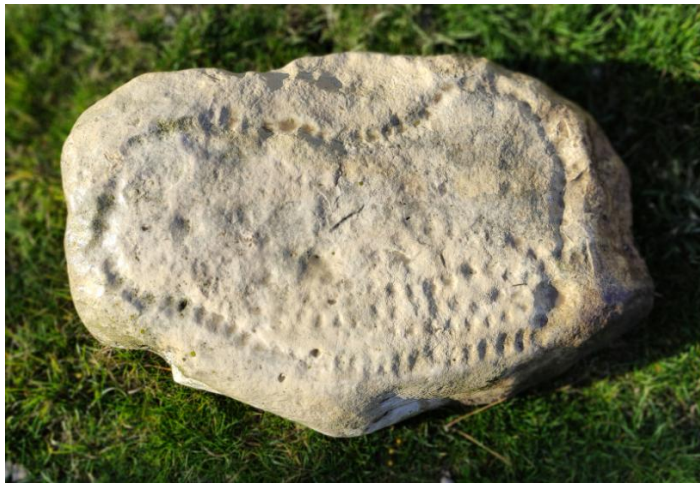

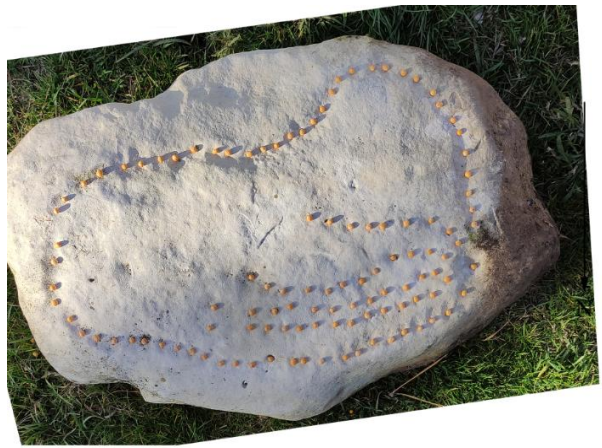

**Figure 8.** Kurgan grave field Prolom II, kurgan 1, slab with cup marks: *a* – photograph of a vertically installed Belogorsk slab on the territory of the Historical and Archaeological Museum-Reserve "Scythian Naples"; *b* – photograph of a slab with a wetted surface in a horizontal position; *c* – photograph of a slab with peas placed in the same type cup marks of the complex.

The Belogorsk slab was attributed to the Srubnaya culture and approximately dated, similarly to the discovered Srubnaya slabs with cup marks, XV-XII centuries BC. Therefore, calculations of the hour markers of the analemmatic sundial for the Belogorsk slab were preliminary made by us for 1500 BC[3]. As geographical coordinates for calculations, the coordinates of the place where the slab was found were taken: 45°07'N, 34°45'E.

Calculations were carried out according to formulas 1-6:

$$M = \frac{m}{\sin \varphi},\tag{1}$$

$$x = M \cdot \sin H,\tag{2}$$

$$y = M \cdot \sin \varphi \cdot \cos H,\tag{3}$$

$$Z_{ws} = M \cdot tg\delta_{ws} \cdot \cos\varphi,\tag{4}$$

$$Z_{ss} = M \cdot tg\delta_{ss} \cdot \cos\varphi,\tag{5}$$

$$H^{/} = arctg\left(tgH/\sin\varphi\right), \text{ при } t \in [6; 18]\tag{6}$$

$$H^{/} = arctg\left(tgH/\sin\varphi\right)\text{-}180°, \text{ при } t \in [0; 6[$$

$$H^{/} = arctg\left(tgH/\sin\varphi\right)\text{+}180°, \text{ при } t \in ]18; 24],$$

$$\text{где } H = 15° \cdot \left(t-12\right),$$

where $x$ – the coordinate of a point along the $X$ axis for an analemmatic sundial, $y$ – the coordinate of a point along the $Y$ axis for an analemmatic sundial, $m$ – the semi-minor axis of the ellipse, $M$ – the semi-major axis of the ellipse, $\varphi$ – the latitude of the area, $t$ – the true local solar time, $H$ – the hour angle of the Sun, $H'$ – the angle between the noon line and the hour line on the clock relative to the center of coordinates (center of the ellipse), $\delta_{ws}=-\varepsilon$ – declination of the Sun on the day of the winter solstice, $\delta_{ss}=\varepsilon$ – declination of the Sun on the day of the summer solstice, $y=Z_{ws}$ – on the day of the winter solstice, $y=Z_{ss}$ – on the day of the summer solstice (Fig. 9). On the days of the equinox $\delta_{eq}=0$.

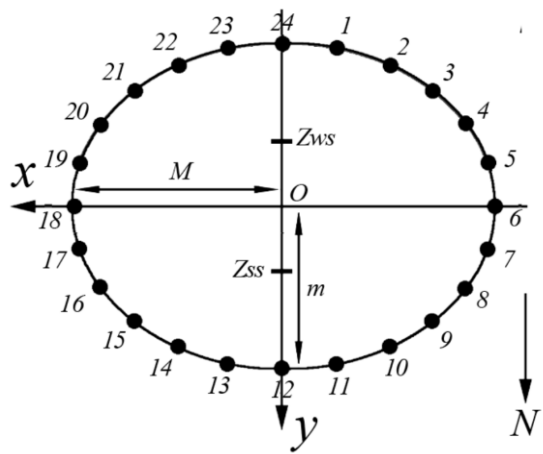

**Figure 9.** Coordinate plane with hour markers from 6 to 18 o'clock. M is the semi-major axis of the ellipse, $m$ is the semi-minor axis of the ellipse, $O$ is the center of the ellipse, $O_{ws}$ is the position of the gnomon on the winter solstice for analemmatic clock, $O_{ss}$ is the position of the gnomon on the summer solstice for the analemmatic clock. N - the direction to the true North.

---

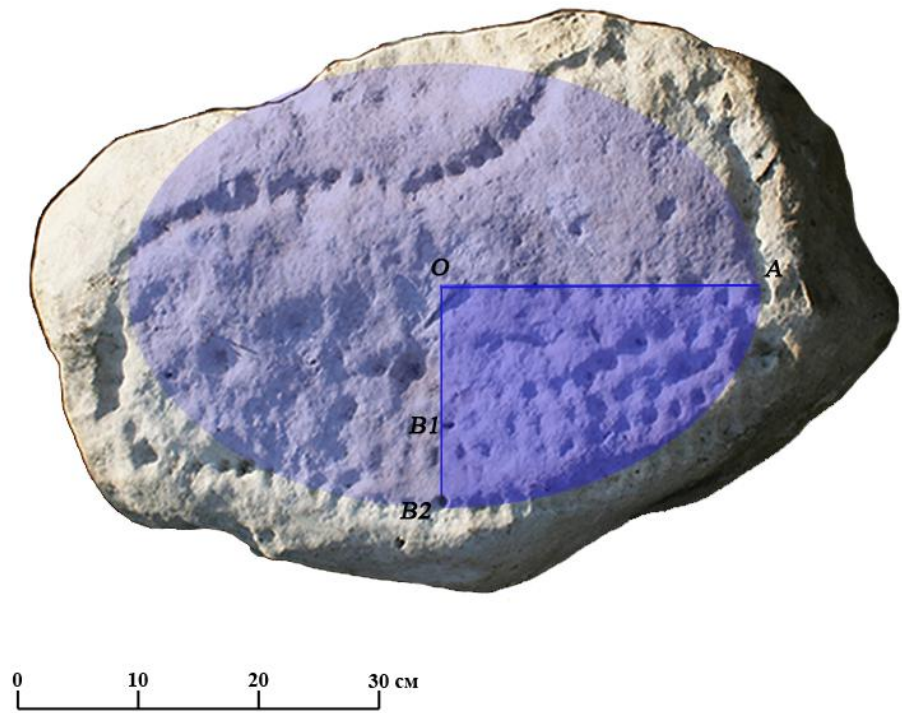

**Figure 10.** The Belogorsk Slab with an ellipse of the hour markers of the analemmatic clock (highlighted in blue), whose major semiaxis is segment OA, and the minor semiaxis is segment OB2.

The segment OA, consisting of dimples linearly located in the central part of the slab, was taken as the major semiaxis M (Fig. 10). As a minor semi-axis, m is the segment OB2, which passes through two cup marks slightly different from the others, corresponding to points B1 and B2. They have a slightly smaller size, but greater depth, are located close to the middle of the slab and set a line perpendicular to segment OA.

The measured length of segment OB2 is the semi-minor axis m≈18 cm, and segment OA is the semi-major axis M≈26 cm. The value of the semi-major axis M=25.4 cm calculated by formula 1 for the latitude of the slab detection, which is close to the measured value M≈26 cm. Such a coincidence of the results of measurements and calculations is one of the confirmations that the Belogorsk slab is indeed an analemmatic sundial.

To mark the analemmatic sundial, the value of the displacement of the gnomon along the Y axis on the day of the winter and summer solstices is also used. It was calculated by formulas 4 and 5 and is equal to Zws≈-8 (cm) and Zss≈8 (cm), respectively. That is, to correctly measure time using an analemmatic sundial, the gnomon had to be shifted on the day of the summer solstice by ≈8 cm to the north of the coordinate center (point O), and on the day of the winter solstice by ≈8 cm to the south of the coordinate center (Fig. 9). At point O, he was supposed to be on the days of the equinox.

The results of calculations using formulas 2 and 3 of the coordinates x and y of the hour marks of the analemmatic clock for the latitude of the slab detection are given in Table 1 and are shown in Figure 11.

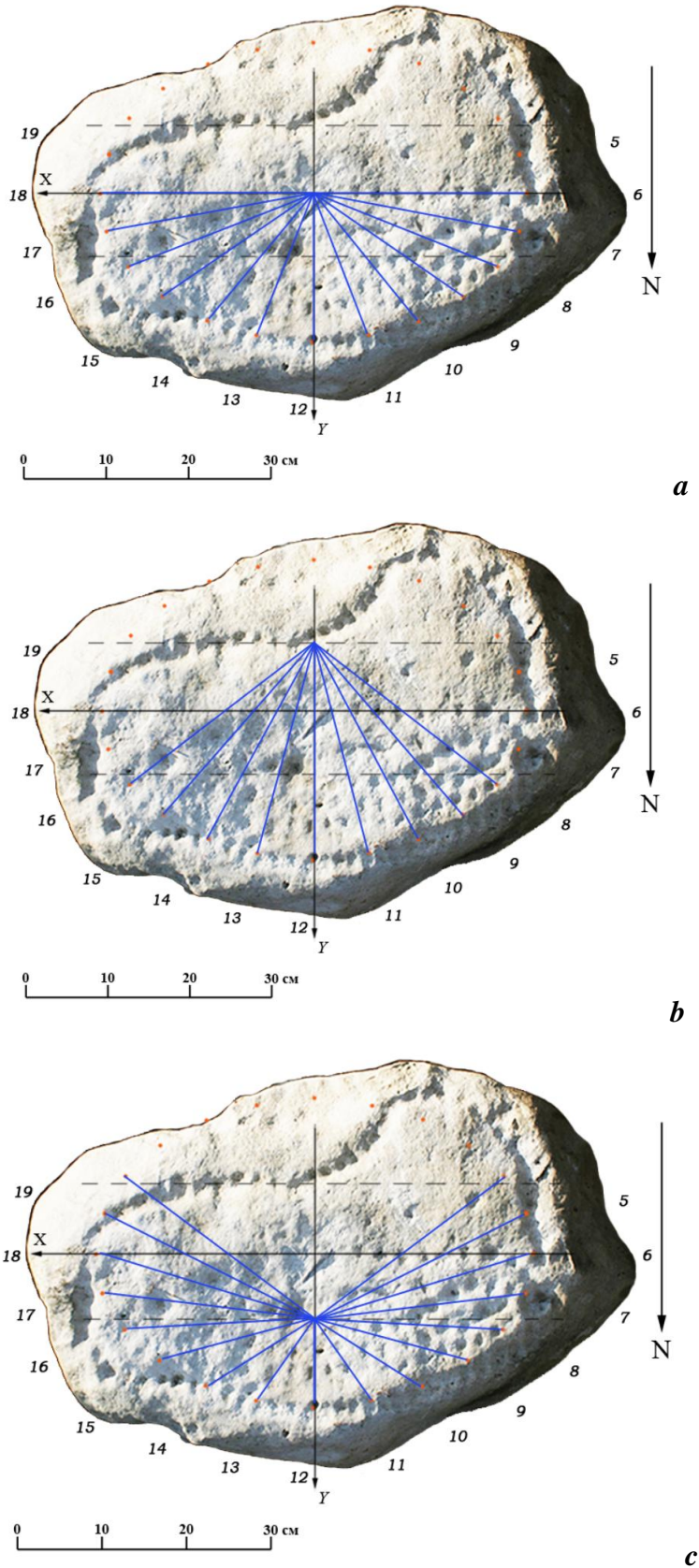

**Figure 11.** Hour marks of an analemmatic sundial projected onto a photograph of the surface of the Belogorsk slab: **_a_** – hour lines on the equinoxes, **_b_** – hour lines on the winter solstice, **_c_** – lhour lines on the summer solstice. Hour markers are applied in orange and signed with the numbers of the corresponding hour (from 5 to 19 hours). N – the direction to true north. The dotted line marks the distances by which the gnomon must be shifted during the summer and winter solstices.

**Table 1.** Calculated coordinates of hour markers of the analemmatic sundial for latitude 45°06'37" N: H – the hour angle of the Sun, H' – the angle between the noon line and the hour line on the sundial, t – mean solar time, x – the coordinate of the hour markers along the X axis, y – the y-coordinate of the clock marks.

| | t, (hour) | | | | | | | | | | | | |
|---|---|---|---|---|---|---|---|---|---|---|---|---|---|
| | 0 | 1 | 2 | 3 | 4 | 5 | 6 | 7 | 8 | 9 | 10 | 11 | 12 |
| H, (⁰) | -180,0 | -165,0 | -150,0 | -135,0 | -120,0 | -105,0 | -90,0 | -75,0 | -60,0 | -45,0 | -30,0 | -15,0 | 0,0 |
| H', (⁰) | -180,0 | -159,2 | -140,8 | -125,2 | -112,2 | -100,7 | -90,0 | -79,3 | -67,8 | -54,8 | -39,2 | -20,8 | 0,0 |
| x, (cm) | 0,0 | -6,6 | -12,7 | -18,0 | -22,0 | -24,6 | -25,4 | -24,6 | -22,0 | -18,0 | -12,7 | -6,6 | 0,0 |
| y, (cm) | -18,0 | -17,4 | -15,6 | -12,7 | -9,0 | -4,7 | 0,0 | 4,7 | 9,0 | 12,7 | 15,6 | 17,4 | 18,0 |

| | t, (hour) | | | | | | | | | | | | |
|---|---|---|---|---|---|---|---|---|---|---|---|---|---|
| | 12 | 13 | 14 | 15 | 16 | 17 | 18 | 19 | 20 | 21 | 22 | 23 | 24 |
| H, (⁰) | 0,0 | 15,0 | 30,0 | 45,0 | 60,0 | 75,0 | 90,0 | 105,0 | 120,0 | 135,0 | 150,0 | 165,0 | 180,0 |
| H', (⁰) | 0,0 | 20,8 | 39,2 | 54,8 | 67,8 | 79,3 | 90,0 | 100,7 | 112,2 | 125,2 | 140,8 | 159,2 | 180,0 |
| x, (cm) | 0,0 | 6,6 | 12,7 | 18,0 | 22,0 | 24,6 | 25,4 | 24,6 | 22,0 | 18,0 | 12,7 | 6,6 | 0,0 |
| y, (cm) | 18,0 | 17,4 | 15,6 | 12,7 | 9,0 | 4,7 | 0,0 | -4,7 | -9,0 | -12,7 | -15,6 | -17,4 | -18,0 |

The results obtained revealed some interesting regularities in the arrangement of dimples, but did not fully explain all its features, first of all, rows of dimples in the northern part of the slab. Therefore, a hypothesis was put forward about the possible connection of these series with the hour lines on other days of the year (not on the days of the equinoxes and solstices) and a decision was made to carry out calculations to construct the analemma.

To accurately measure time in a solar analemmatic clock, the gnomon must be moved not just along the Y axis (North-South axis), but along the analemma, which looks like a figure eight and changes depending on the epoch. The analemma reflects the position of the Sun at the same mean solar time of the day throughout the year, which occurs due to the mismatch between the true solar time and the mean solar time (Fig. 12).

An analemma for a sundial could have been built empirically in antiquity, with the simultaneous measurement of time with the help of sundial and water clock. The sundial shows true solar time. Mean solar time can be measured with a water clock. In order to achieve the coincidence of the readings of the solar analemmatic clock with the readings of the water clock, it was necessary that the gnomon move throughout the year in accordance with the solar analemma. By the way, clepsydras (water clocks) of the Bronze Age have already been discovered in the Northern Black Sea region (Vodolazhskaya, Usachuk, Nevsky, 2015a; Vodolazhskaya, Novichikhin, Nevsky, 2021).

The unevenness of the daily motion of the Sun is due to the ellipticity of the Earth's orbit around the Sun and the inclination of the Earth's axis to the plane of the ecliptic. The duration of an average solar day is the average value of the duration of a true solar day per year. The difference between true solar time and mean solar time at the same moment is called the equation of time4 and is calculated using formula 7:

$$\eta = T_s - T_m \ , \qquad\qquad (7)$$

---

[4] Sometimes an inverted time equation is used, which is equal to the difference between mean time and true solar time.

where $\eta$ - the equation of time (minutes), $T_s$ - true solar time, $T_m$ - mean solar time. If the value of the equation of time is negative, then the true solar time lags behind the mean solar time, determined, for example, by the water clock, and if it is positive, then it is ahead.

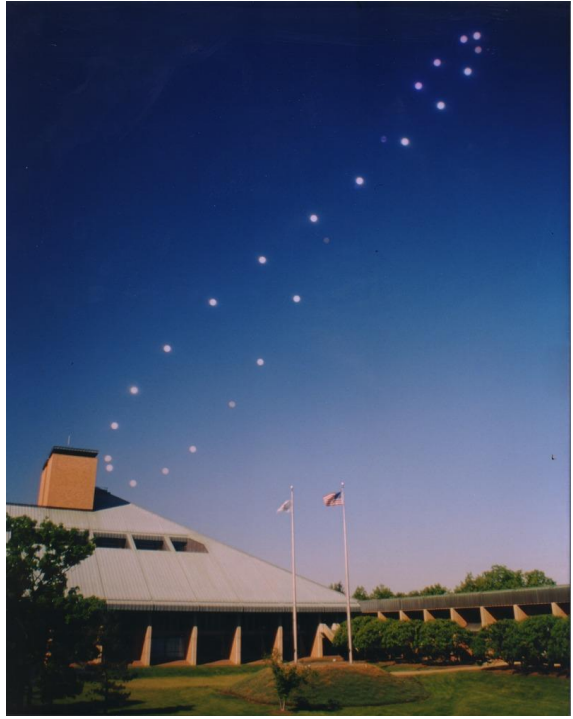

**Figure 12.** Midday solar analemma[5].

Calculations of the equation of time were carried out by us using the astronomical program HORIZONS System[6]. The calculation results for each month, starting from the day of the vernal equinox, are presented in Table 2[7].

The movement of the gnomon in the analemmatic sundial along the $Y$ axis was calculated using formula 8, and along the $X$ axis it was approximately calculated using formula 9, which we obtained:

$$Z_y = M \cdot tg\,\delta \cdot \cos\varphi, \tag{8}$$

$$Z_x = \eta \cdot \frac{M \cdot \sin\left(15^\circ\right)}{60}, \tag{9}$$

where $Z_y$ - the displacement of the gnomon along the $Y$ axis, $Z_x$ - the displacement of the gnomon along the $X$ axis, $M$ - the semi-major axis of the ellipse, $\delta$ - the declination of the Sun, $\varphi$ - the latitude of the area, $\eta$ - the equation of time.

The dates of the equinoxes, solstices, and the times of sunrise and sunset at the beginning of each month were calculated using the astronomical program RedShift 7.

The calculation results are presented in Tables 3, 4. The drawing of the analemma is shown in Figure 13.

**Table 2.** Coordinates of analemma points for 1500 BC: $Z_y$ - displacement of the gnomon along the $Y$ axis, $Z_x$ - displacement of the gnomon along the $X$ axis, $\delta$ - declination of the Sun, $\eta$ - equation of time.

---

| Point number on the analemma | Date | 1500 BC | | | |
|---|---|---|---|---|---|
| | | $\delta$, ° | $\eta$, min | $Z_x$, cm | $Z_y$, cm |
| 1 | April (vernal equinox) 3 April 1500 BC | -0,05 | -7 | -0,8 | 0,0 |
| 2 | May 3 May 1500 BC | 11,13 | 5 | 0,5 | 3,5 |
| 3 | June 3 June 1500 BC | 20,07 | 10 | 1,1 | 6,6 |
| 4 | July 3 July 1500 BC | 23,83 | 6 | 0,7 | 7,9 |
| 5 | August 3 August 1500 BC | 21,20 | -2 | -0,2 | 7,0 |
| 6 | September 3 September 1500 BC | 12,67 | -2 | -0,2 | 4,0 |
| 7 | October 3 October 1500 BC | 1,12 | 4 | 0,4 | 0,4 |
| 8 | November 3 November 1500 BC | -11,22 | 9 | 1,0 | -3,6 |
| 9 | December 3 December 1500 BC | -20,37 | 6 | 0,7 | -6,7 |
| 10 | January 3 January 1500 BC | -23,82 | -8 | -0,9 | -7,9 |
| 11 | February 3 February 1500 BC | -19,98 | -19 | -2,1 | -6,5 |
| 12 | March 3 March 1500 BC | -11,72 | -18 | -2,0 | -3,7 |

The equation of time in 1500 BC is equal to zero on April 19-20 and July 24-27 (Fig. 13). On these days, the time shown by the sundial and water clock coincided and the movement of the gnomon along the X axis was not required.

**Table 3.** Coordinates of the analemma points on the days of the solstices and equinoxes in 1500 BC: $\delta$ - declination of the Sun, $\eta$ - equation of time, $Z_x$ - displacement of the gnomon along the $X$ axis, $Z_y$ - displacement of the gnomon along the $Y$ axis.

| Designation points | Astronomical phenomenon | Date | $\delta$ | $\eta$, min | $Z_x$, cm | $Z_y$, cm |
|---|---|---|---|---|---|---|
| VE | Vernal equinox | 3 April | 0° | -7 | -0,8 | 0,0 |
| SS | Summer solstice | 6 July | 23,90° | 5 | 0,5 | 8,0 |
| AE | Aautumnal equinox | 6 October | 0° | 5 | 0,5 | 0,0 |
| WS | Winter solstice | 2 January | -23,90° | -7 | -0,8 | -8,0 |

**Table 4.** The times of sunrise and sunset at the beginning of each month[8] after the equinox, as well as on the days of the solstices and the autumnal equinox in 1500 BC. for geographic coordinates 45°04′ N, 34°36′ E

---

[8] This refers to the synodic lunar month - the time interval between two successive identical phases of the moon (for example, new moons or full moons), approximately equal to 30 days.

| Designation points | Astronomical phenomenon | Date | Sunrise | Sunset |
|---|---|---|---|---|
| VE | Vernal equinox | April 3 | 5:58 | 18:03 |
| 1 | 1st month (after the Vernal equinox) | May 3 | 5:12 | 18:49 |
| 2 | 2nd month | June 3 | 4:31 | 19:30 |
| 3 | 3rd month | July 3 | 4:11 | 19:49 |
| SS | Summer solstice | July 6 | 4:11 | 19:49 |
| 4 | 4th month | August 3 | 4:24 | 19:35 |
| 5 | 5th month | September 3 | 5:05 | 18:55 |
| 6 | 6th month | October 3 | 5:52 | 18:07 |
| AE | Autumnal equinox | October 6 | 5:57 | 18:02 |
| 7 | 7th month | November 3 | 6:47 | 17:12 |
| 8 | 8th month | December 3 | 7:27 | 16:33 |
| WS | Winter solstice | January 2 | 7:42 | 16:19 |
| 9 | 9th month | January 3 | 7:41 | 16:18 |
| 10 | 10th month | February 3 | 7:22 | 16:38 |
| 11 | 11th month | March 3 | 6:45 | 17:15 |

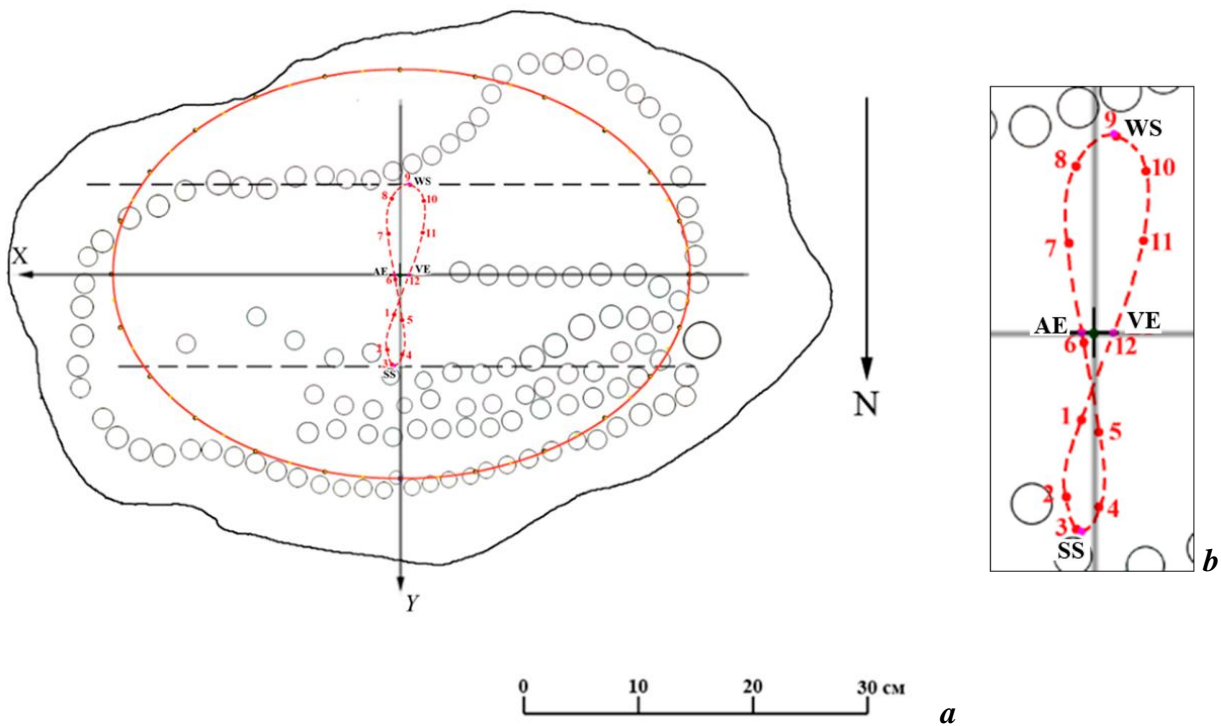

**Figure 13.** Plan-scheme of the Belogorsk slab: *a* – surface of the slab with an applied analemma for a moving gnomon and an ellipse of hour markers; *b* – an enlarged image of the analemma for a mobile gnomon. WS – the point at which you need to install the gnomon on the day of the winter solstice, SS – on the summer solstice, AE – on the autumnal equinox, VE – on the vernal equinox. The numbers on the analemma show the points at which the gnomon must be set in the corresponding number of months after the vernal equinox.

Several rows of cup marks could have arisen if an attempt was made to make the gnomon stationary and move the "dial", replacing the gnomon with several "dials", each of which would correspond to its own month.

The analemma for cup marks should be a mirror image horizontally and vertically of the analemma for the gnomon. It was built in a graphics editor using a geometric transformation (Fig. 14).

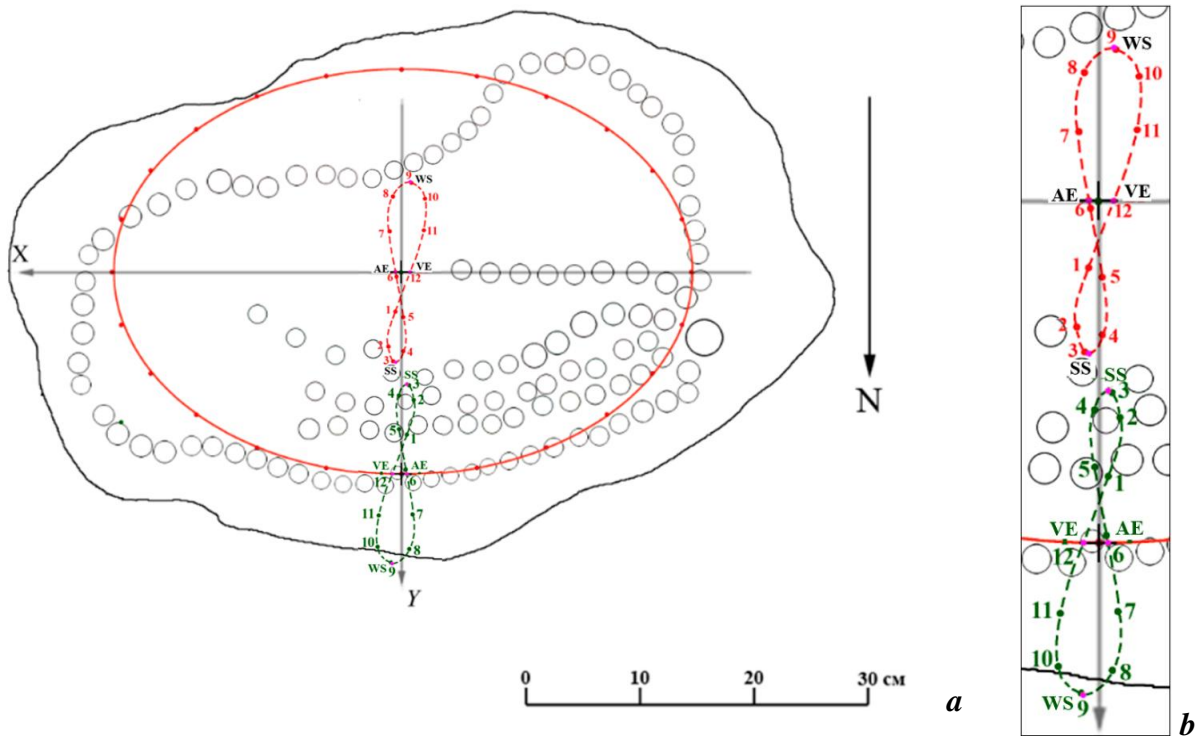

**Figure 14.** Plan-scheme of the Belogorsk slab: ***a*** – the surface of the slab with an analemma for a moving gnomon with the corresponding ellipse of hour marks (highlighted in red) and an analemma for cup marks corresponding to 12 hours during the year every month (highlighted in green); ***b*** – enlarged image of both analemmas. WS – the point at which you need to install a gnomon (red analemma) or place a cup mark (green analemma) on the day of the winter solstice, SS – on the summer solstice, AE - on the autumnal equinox, VE – on the vernal equinox. The numbers on the analemma show the points where you need to install the gnomon (red analemma) or place the cup marks 12 hours after the corresponding number of months after the vernal equinox (green analemma).

The points corresponding to the required location of the cup marks on the days of the equinoxes, solstices and every month from the day of the vernal equinox are marked on the analemma (Fig. 4b). In the case of a stationary vertically standing gnomon, each hour mark will have its own analemma, identical to the analemma of the 12 o'clock mark (Fig. 15a). However, analemmas for only whole hours do not explain all the dimples in the northern part of the slab. Only the addition of analemmas for every half hour allows successful interpretation of most of the cup marks (Fig. 15b).

Thus, for each month, its own ellipse of hour markers is set, along which the cup marks associated with it are located (Fig. 16). Based on this principle, one can try to reconstruct in detail the entire technology of measuring time using the Belogorsk slab.

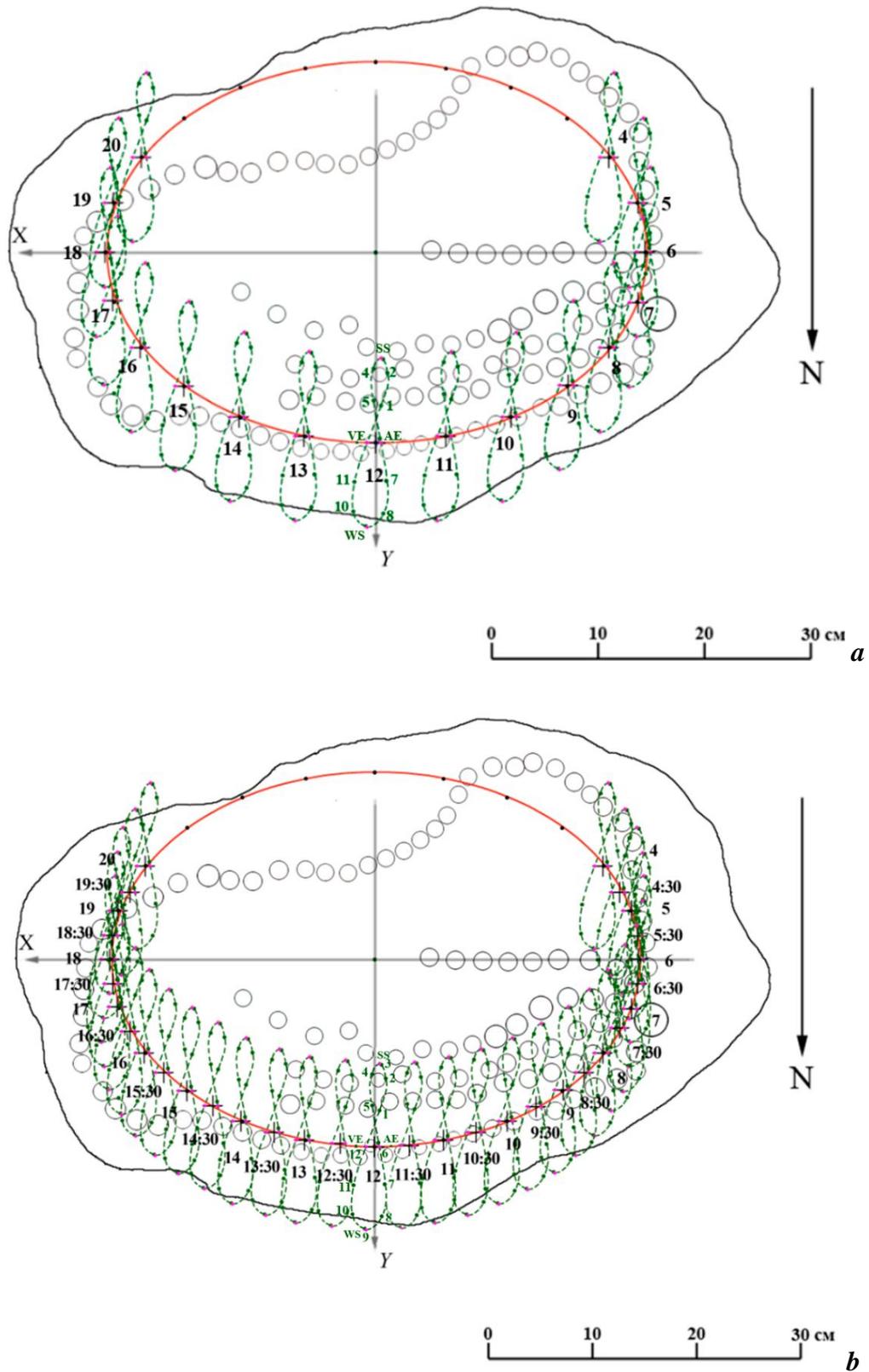

**Figure 15.** Plan-scheme of the Belogorsk slab: *a* – surface of the slab with analemmas of hour marks for whole hours; *b* – the surface of the slab with analemmas of hour marks every half an hour.

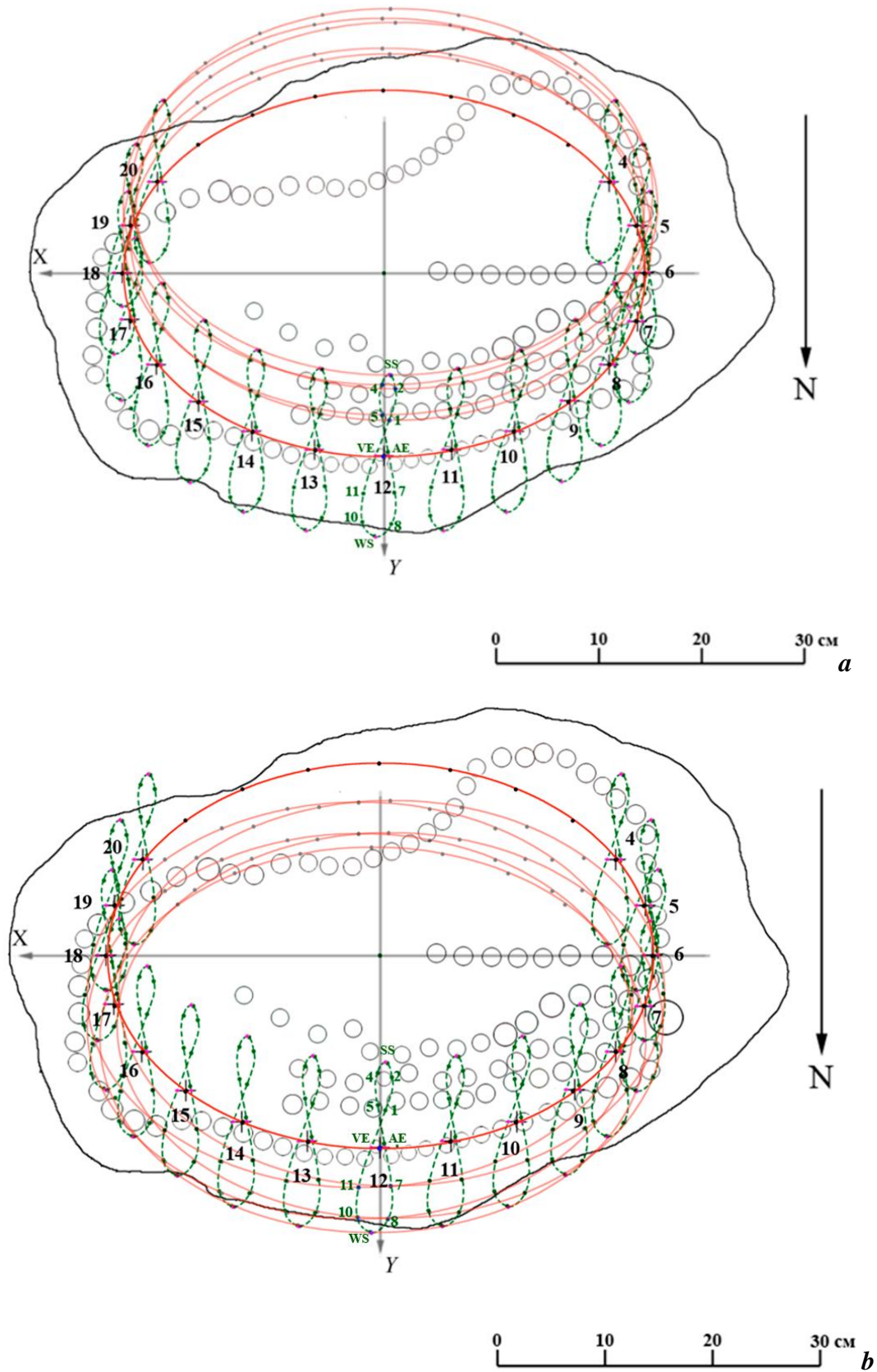

**Figure 16.** Plan-scheme of the Belogorsk slab with analemmas and ellipses of hour markers for points on analemmas corresponding to months: ***a*** – from vernal to autumnal equinox; ***b*** – from autumn to vernal equinox.

The first row of cup marks on the side of the northern edge of the slab refers to the equinox. Cup marks, close to the hour markers of the vernal equinox (VE) on the analemmas for cup marks, are marked in yellow in Figure 17.

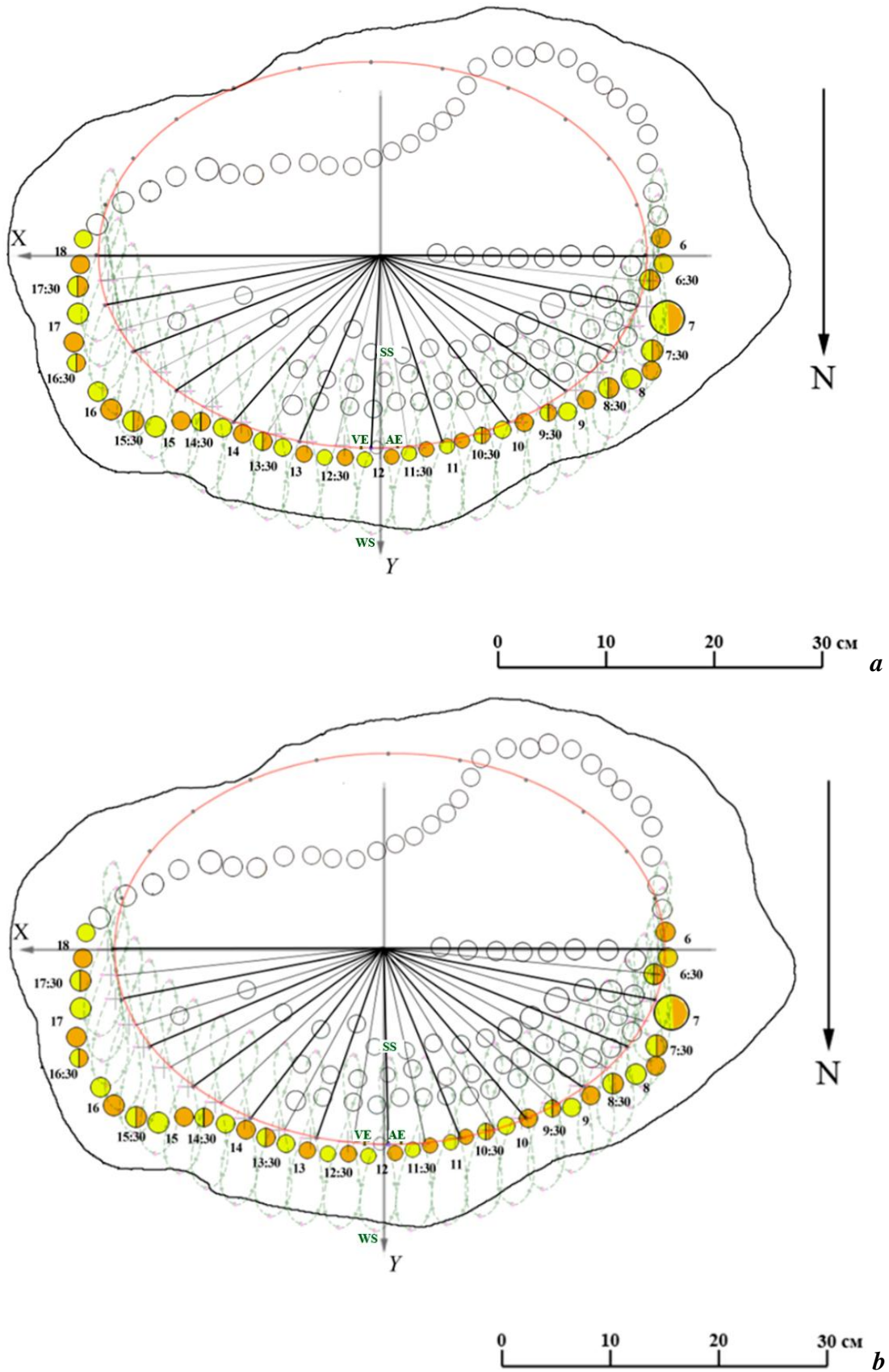

**Figure 17.** Plan-scheme of the Belogorsk slab. Hour lines and corresponding cup marks for measuring time at the equinox: **a** – vernel equinox; **b** – autumnal equinox. Cup marks for measuring time only in the vicinity of the autumnal equinox are colored orange, only in the vicinity of the vernal equinox – in yellow, for measuring time in the vicinity of both equinoxes – in both colors.

They could be used to measure time in a period close to the vernal equinox. Cup marks that turned out to be close to the hour marks of the autumnal equinox (AE) are highlighted in orange on the analemmas for cup marks in Figure 17. They were used to measure time in a period close to the autumnal equinox. The corresponding hours are signed in the figure under the cup marks. The cup marks, the halves of which are colored both yellow and orange, were used to measure the time at both equinoxes.

In the arrangement of cup marks for measuring time at different equinoxes, a simple pattern is visible, which, most likely, was used by the creator of this sundial. First (clockwise) there is a cup mark for a whole hour on the autumn equinox (orange), then for the same hour on the vernal equinox (yellow), and then a cup mark common to both equinoxes (two-color), corresponding to half of the next hour.

This principle works almost all the way through the first row of cup marks. An exception is the region near 12 o'clock, where 11:30 and 12:30 correspond not to one two-color cup mark, but to two cup marks with the same alternation of colors as in the case of whole hours. This is most likely due to the fact that the hour angles near the 12 o'clock mark have a maximum value, and the cup marks on the Belogorsk slab are applied with approximately the same density along the entire first row.

The second exception is the cup mark corresponding to 7 o'clock. Instead of two small cup marks on the surface of the slab, one large one is knocked out. It may have been a control cup mark that served as a starting point. Knowing that it corresponds to seven hours, and knowing the regularity of the alternation of the cup marks of the autumn and vernal equinoxes, it was easy to determine the hour that a particular cup mark corresponds to.

It should also be noted that the slight deviation of first row of dimples from the ellipse in the northwestern part of the slab and a slightly more noticeable deviation in the northeastern part do not seem to be accidental. Cup marks corresponding to the ellipse of hour markers around midday, for the day of the winter solstice, are missing on the Belogorsk slab. But deviations in the first row of cup marks are observed precisely on those cup marks on which the visible time of day begins and ends at the winter solstice. On the winter solstice, the Sun rises at 7:42 and the Sun sets at 16:19 (Table 5). The nearest cup marks: 7:30 and 16:30 practically mark these deviations (Fig. 18a). That is, the 7:30, 16, and 16:30 cup marks correspond quite closely to the hour markers of the winter solstice marker ellipse. That is, it seems that on the day of the winter solstice and around it, it was important to determine the time only at sunrise and before sunset.

One month before (the 8th month after the vernal equinox (see Fig. 14b)) and one month after the winter solstice (the 10th month after the vernal equinox), the cup marks turn out in the area of deviation of the first row of cup marks from the ellipse of the hour markers as well. Sunrise at 7:27 and 7:41, and sunset at 16:33 and 16:18 one month before and one month after the winter solstice, respectively (Table 5). The nearest cup marks: 7:30 and 16:30 coincide with the hour marks for the winter solstice on the analemmas for cup marks and are close to the corresponding hour marks of the mark ellipses of the 8th and 10th months after the vernal equinox (Fig. 18b, 18c). That is, in the period from one month before and after the winter solstice, it was important to determine the time only at times near sunrise and sunset. These are the shortest and coldest days of the year, which, as a rule, are also cloudy, i.e. measuring time with a sundial becomes unlikely these days.

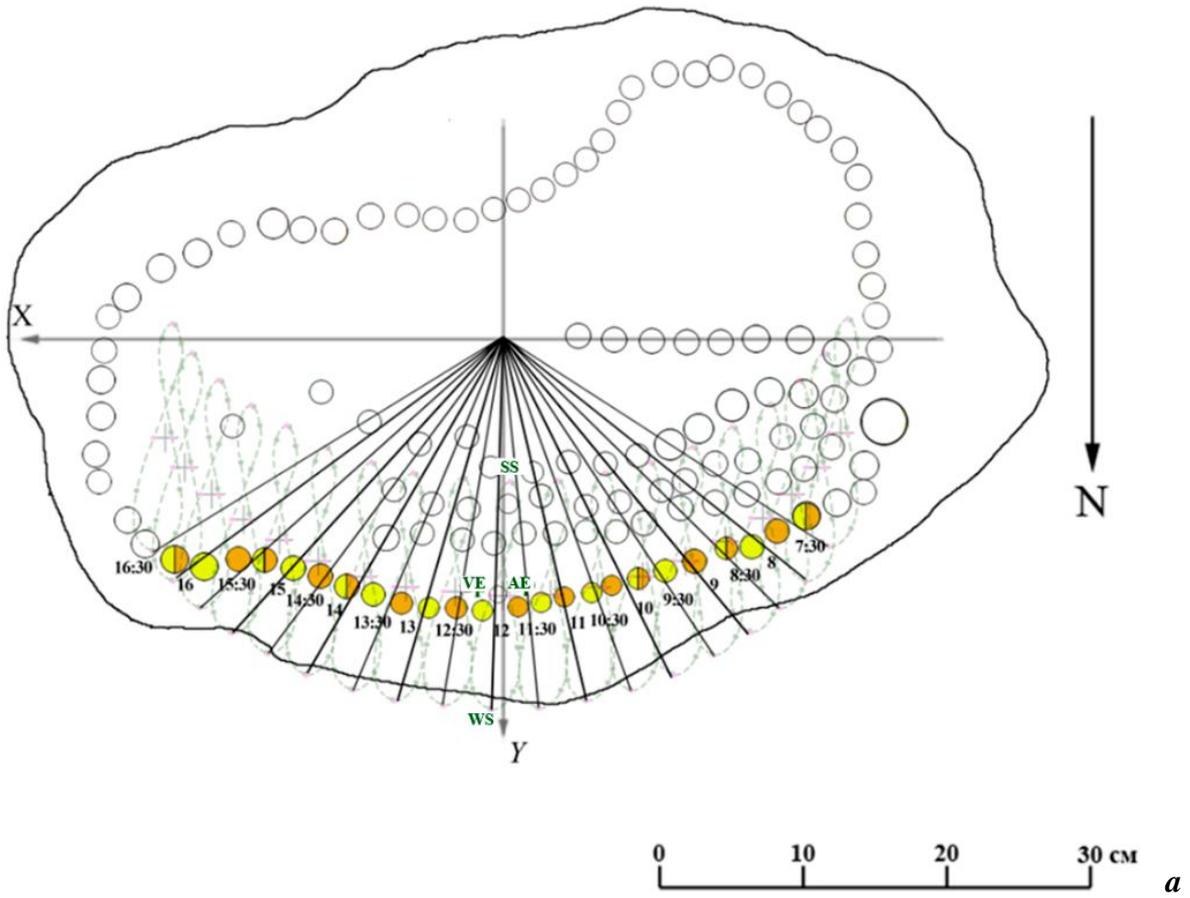

*a*

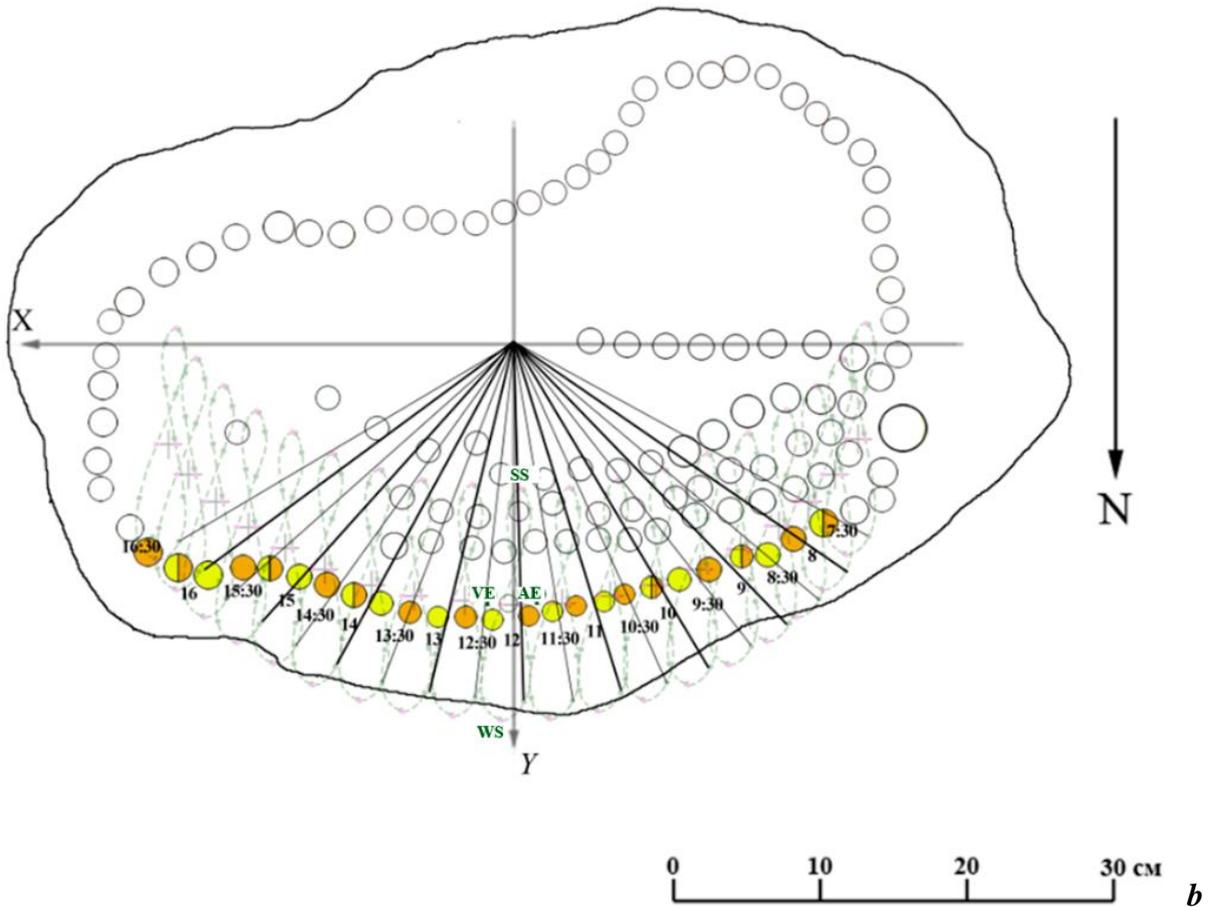

*b*

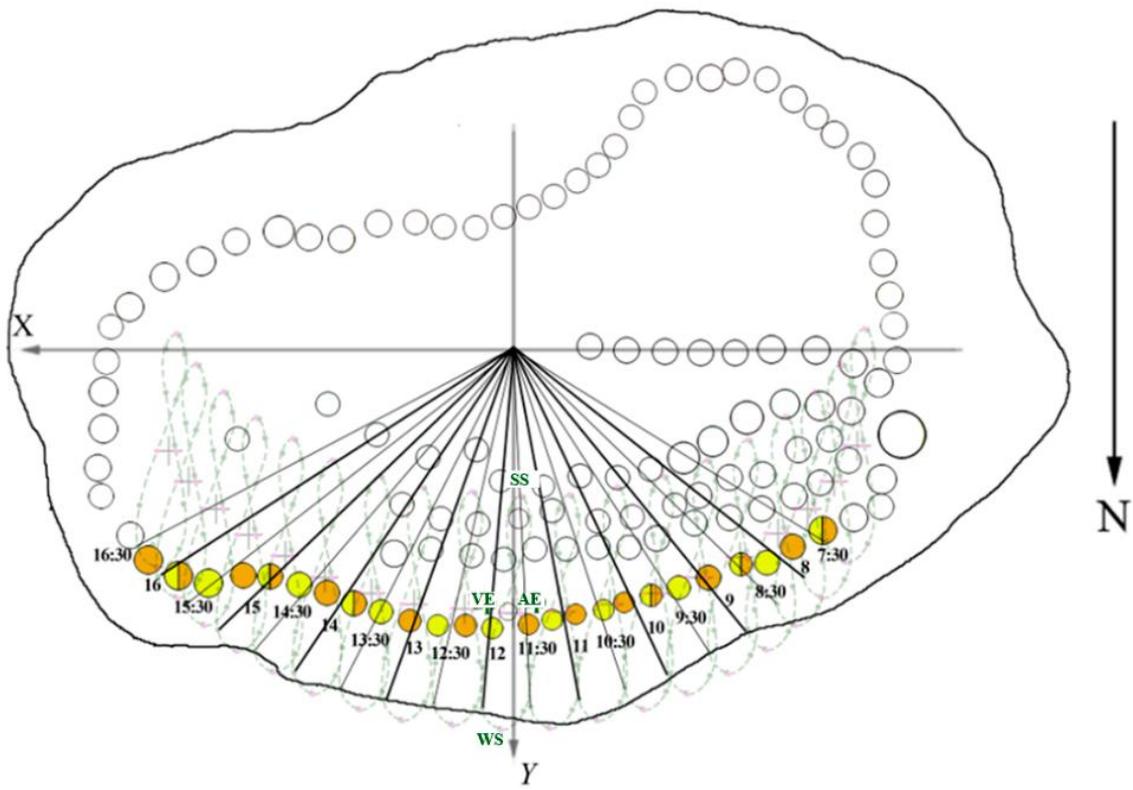

*c*

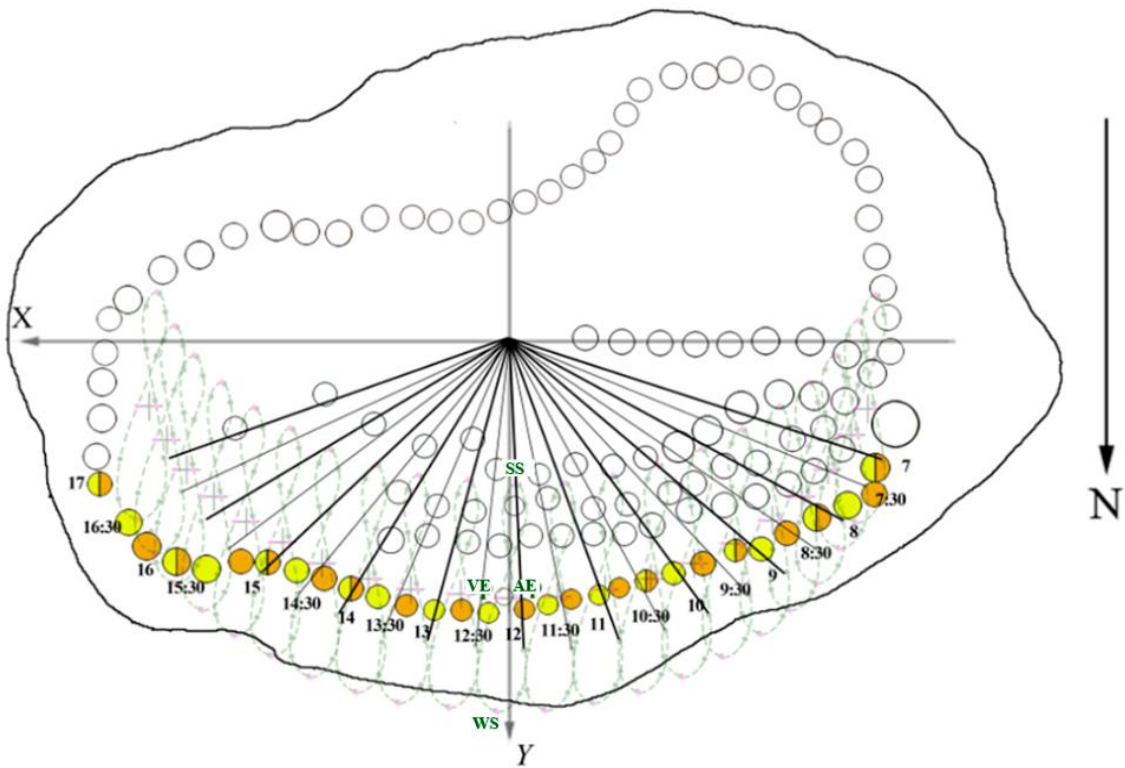

*d*

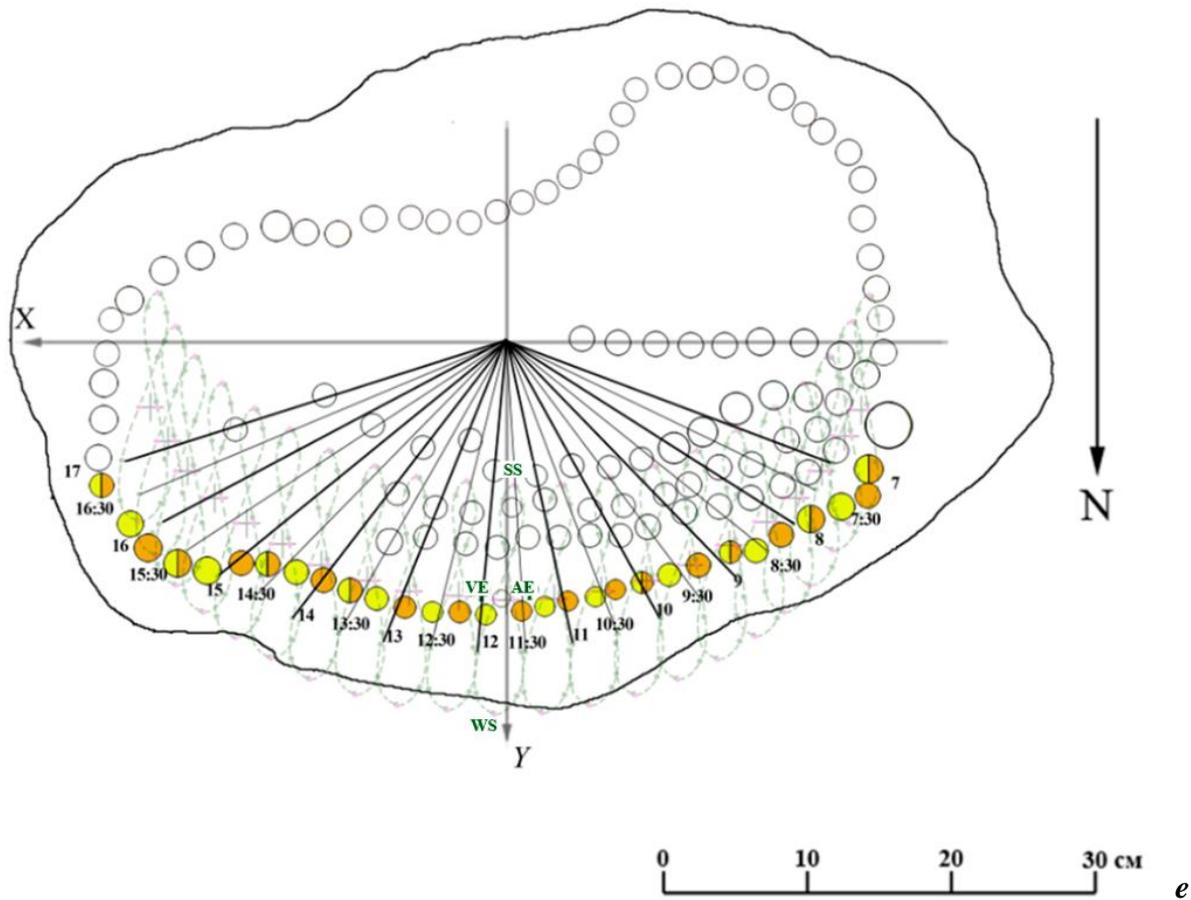

**Figure 18.** Plan-scheme of the Belogorskaya slab with a schematic drawing of cup marks. Hour lines and marks: ***a*** – for the winter solstice (WS) (9 months after the vernal equinox (VE)); ***b*** – one month before the WS (8 months after the VE); ***c*** – one month after the WS (10 months after the VE); ***d*** – two months before the WS (7 months after the VE), ***e*** – two months after the WS (11 months after the VE). Cup marks for measuring time only in the vicinity of the autumnal equinox are colored orange, only in the vicinity of the vernal equinox - in yellow, for measuring time in the vicinity of both equinoxes - in both colors.

Two months before (7th month after the vernal equinox) and two months after the winter solstice (11th month after the vernal equinox), the hour marks, on which the apparent time of day begins and ends, also correspond to the area of deviation of the first row of cup marks from ellipse of hour markers. Sunrise occurs at 6:47 and 6:45, and sunset occurs at 17:12 and 17:15 two months before and two months after the winter solstice, respectively (Table 5). The nearest cup mark to the time of sunrise - cup mark 7:00 - is close to the corresponding hour mark of the ellipse of marks of the 7th month after the vernal equinox (Fig. 18d). The nearest cup mark to the time of sunset (cup mark 17:00) is close to the corresponding hour mark of the mark ellipse of the 11th month after the vernal equinox (Fig. 18e).

The rest of the time, the hour marks of the ellipses of the 7th and 11th months already become close to the hour marks of the autumn and vernal equinoxes, respectively. And if desired, the time at 7 and 11 months after the vernal equinox can be approximately measured by the cup marks of the first row using the scheme for the autumn and vernal equinoxes, respectively.

The greater coincidence of the cup marks with the hour marks of the months close to the winter solstice in the areas of deviation of the cup marks of the first row from the hour marks of the equinoxes may indicate the greater importance of knowing the onset of the evening hours in winter, when daylight hours end very early.

The second row of cup marks on the side of the northern edge of the slab refers to the 1st and 5th months after the vernal equinox. Cup marks, that were close to the hour marks of the 1st month on the analemmas for cup marks, are highlighted in blue in Figure 19. According to them, it was possible to measure the time in the period close to the 1st month after the VE. Cup marks that are close to the hour marks of the 5th month are highlighted in green. They were used to measure time in the period close to the 5th month after the VE. The corresponding hours are signed in the figure under the cup marks. Cup marks, the halves of which are colored both blue and green, were used to measure the time in the 1st month and in the 5th month after the VE.

In the location of the cup marks for measuring time in the 1st and 5th months after the VE, a simple pattern is also visible, similar to the pattern associated with the equinox. Starting at 8 o'clock (clockwise), there is a cup mark for the whole hour of the 1st month after the VE (blue), then for the same hour of the 5th month (green), and then - a common to both equinoxes cup mark (two-color) corresponding to half of the next hour. And so on until 13 hours without exception. However, unlike the equinox scheme, all cup marks from the time of sunrise - from about 5 am to 7:30 inclusive should have been used to measure time in both the 1st and 5th months after the WFD. Moreover, in this range, for whole hours and for half hours, one cup mark was placed in accordance with each hour mark.

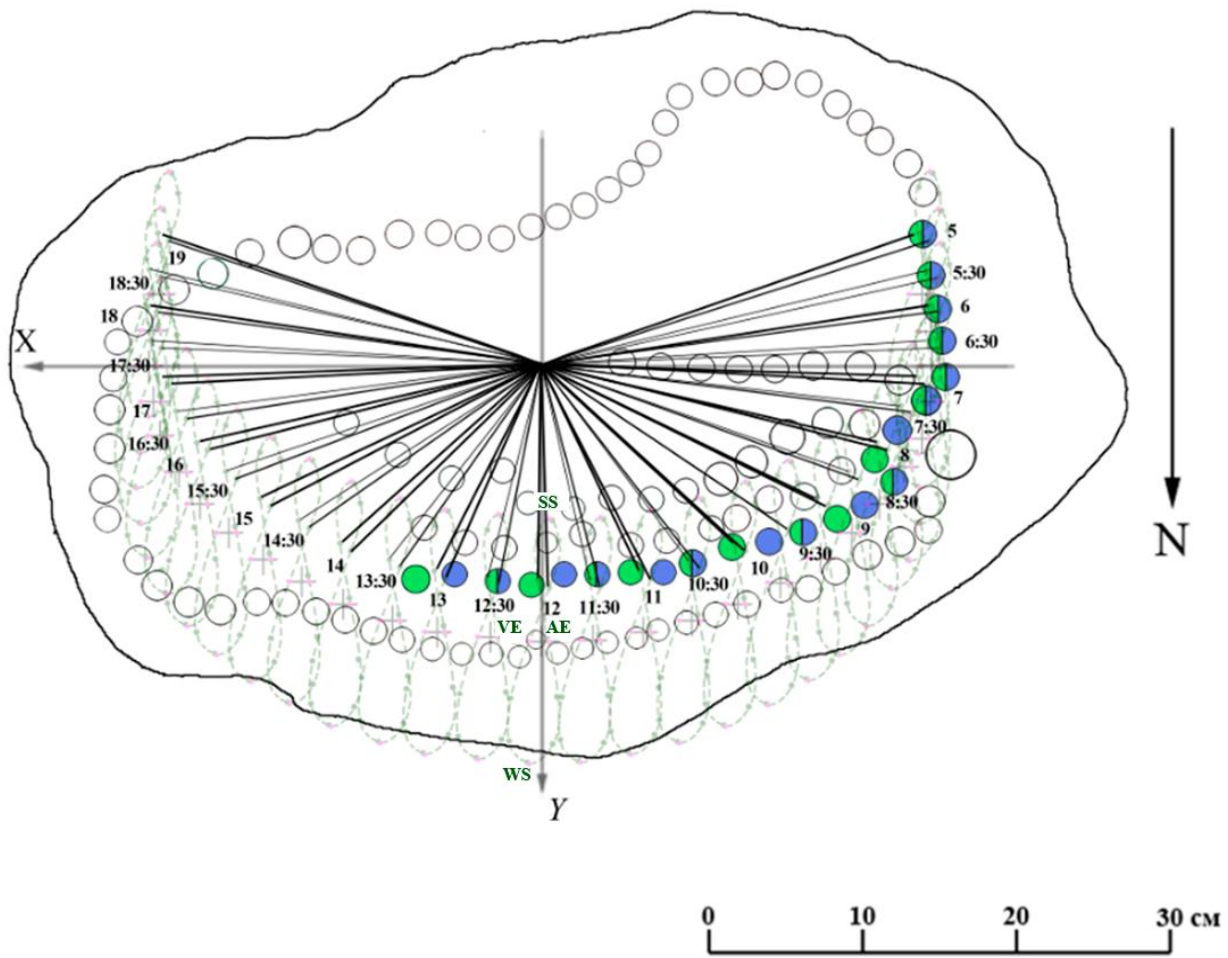

**Figure 19.** Plan-scheme of the Belogorsk slab. Hour lines and corresponding cup marks to measure the time approximately one (highlighted in blue) and five months (highlighted in green) after the vernal equinox. Cup marks for the 1st month after the VE are colored blue, for the 5th month they are green, the cup marks common for both equinoxes are in both colors.

Cup marks corresponding to the hour marks after 13:00, for the 1st and 5th months after the VE and similar to the complex of cup marks in the northwestern part of the slab in terms of application technique, are not on the slab (see Fig. 8a). Perhaps this was due to the irrelevance of measuring the exact time in the afternoon period during the warm season (between the vernal and autumn equinoxes), when most work, primarily agricultural, was carried out in the first half of the day - from dawn to midday, when, especially in summer, it was too hot to work outdoors.

The third row of cup marks on the side of the northern edge of the slab refers to the 2nd and 4th months after the spring equinox. Cup marks, that were close to the hour marks of the 2nd month on the analemmas for cup marks, are highlighted in blue in Figure 20. They could be used to measure time in a period close to the 2nd month after the VE. The cup marks that are close to the hour markers of the 4th month on the analemmas for cup marks are highlighted in pink. They were used to measure time in a period close to the 4th month after the VE. The corresponding hours are signed in the figure under the cup marks. Cup marks, the halves of which are colored in both blue and pink, were used for time measurements in both the 2nd and 4th months after the VE.

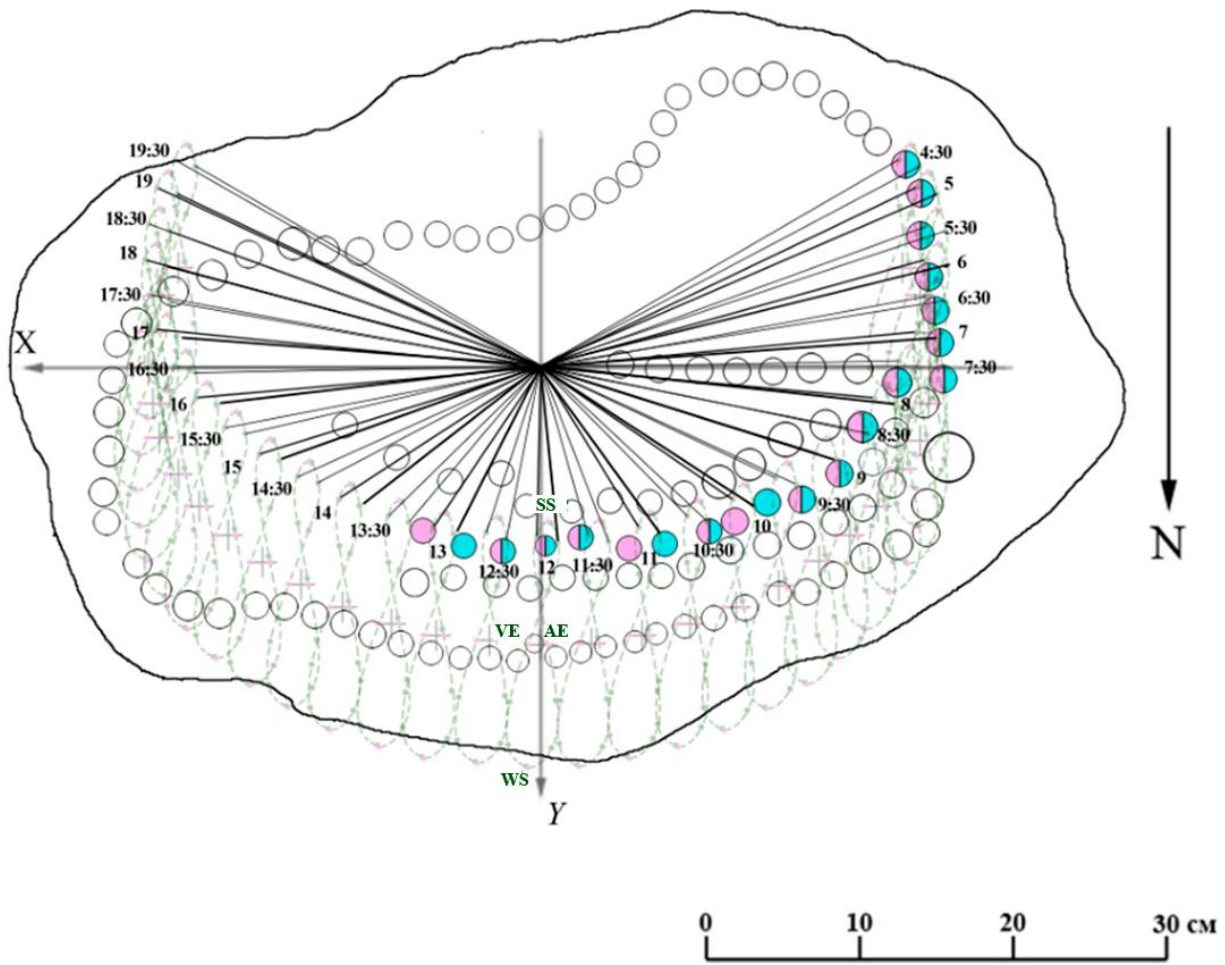

**Figure 20.** Plan-scheme of the Belogorsk slab. Hour lines and corresponding cup marks for measuring time in the period close to two months (blue) and four months (pink) after the vernal equinox. Two-color cup marks are used to measure time in both cases.

In the location of the cup marks for measuring time in the 2nd and 4th months after the VE, there is also a pattern similar to the pattern associated with the equinoxes. Starting at 10 o'clock (clockwise) there is a cup mark for the whole hour of the 2nd month after the VE (blue), and then a cup mark for the same hour of the 4th month (pink). Then there is a cup mark common to both equinoxes (two-color), corresponding to half of the next hour. And so on until 13:00, but with an exception for 12:00. This is one cup mark, and common for both months. However, unlike the scheme of the equinoxes, but similar to the scheme for the 1st and 5th months, all cup marks from the time of sunrise, from about 4:30 in the morning until 9:30 inclusive, should have been used to measure time and for the 2nd and for the 4th month after VE. Moreover, in this range, both for whole hours and for half an hour, one cup mark was put in correspondence with each hour mark.

Cup marks corresponding to the hour marks after 13:00 for the 2nd and 4th months after the WFD, similar in terms of application technique to the complex of cup marks in the northwestern part of the slab, are also not observed on the slab (see Fig. 8a). It is possible that for the same reasons.The fourth row of cup marks on the side of the northern edge of the slab refers to the summer solstice or the 3rd month after the VE. The cup marks that turned out to be close to the hour marks of the whole hours of the 3rd month on the analemma for the cup marks are highlighted in red in Figure 21, and those close to the hour marks of the half hours are highlighted in pale pink. The corresponding hours are signed in the figure under the cup marks. Each whole and each half hour from the moment of sunrise is assigned one cup mark and only 12 hours - two cup marks.

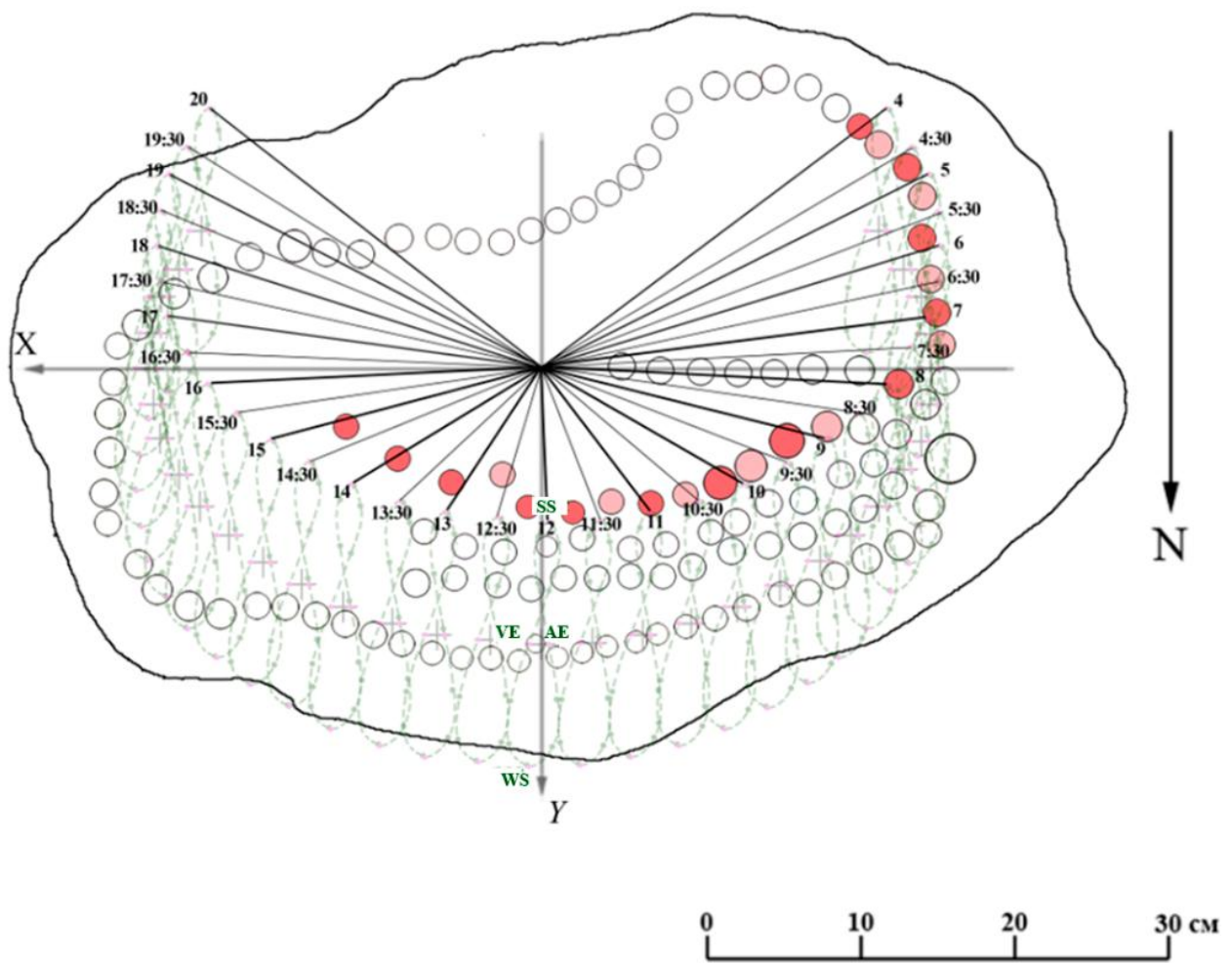

**Figure 21.** Plan-scheme of the Belogorsk slab. Hour lines and corresponding cup marks for measuring time on the days of the summer solstice. Cup marks corresponding to whole hours are highlighted in red, corresponding to halves of an hour in pale pink.

The cup marks corresponding to 13, 14 and 15 o'clock are somewhat different from the rest of the cup marks in the complex and are at some distance from the hour marks, which correspond only approximately. Therefore, it is possible that they are not directly related to the complex of cup marks corresponding to the analemmatic sundial (with analemmas for cup marks), or they played an auxiliary role in the process of marking the slab.

For the completeness of the reconstruction of the technology of measuring time using the Belogorsk slab, it is necessary to clarify the necessary working dimensions of the gnomon. At midday of the summer solstice, its shadow must be no less than the distance from the center of coordinates to the corresponding hour mark of 12 o'clock, which is equal to b≈10 cm. The height of the Sun above the horizon[9] (altitude) at the summer solstice at noon in 1500 BC[10] – $A_{SS}$≈69° at the latitude of detection of the Belogorsk slab. Then the height of the gnomon $a$ will be $a=b*tg(A_{SS})$≈26 cm, i.e. the height of the gnomon will be approximately equal to the length of the semi-major axis of the ellipse of hour markers M. At the equinox, the height of the Sun at noon will be $A_{VE}$≈45°, and the length of the shadow from the gnomon at noon will be ≈26 cm and reach the northern edge of the slab.

In conclusion, I would like to note that the outer row of cup marks - the "dial" in the case of conventional analemmatic clocks should have had the shape of an ellipse. In the case of the Belogorsk slab, it has such a strange shape not by chance. The northern - day part of the "dial" is formed by cup marks corresponding to the hour marks of the equinoxes (Fig. 16). The western - morning part of the "dial" is formed by cup marks corresponding to the hour markers of the day of the summer solstice and the months closest to it (Fig. 16a). The eastern - evening part of the "dial" is formed by cup marks corresponding to the hour markers of the winter solstice and the months closest to it (Fig. 16b). The southern - night part of the "dial", which only reflects the principle of marking the analemmatic sundial, but is not intended for direct measurement of time, serves as a transition from the ellipse of the evening - "winter" hour marks to the ellipse of the morning - "summer" hour marks. For the transition, instead of drawing several rows of cup marks in the southern part of the slab, by analogy with the northern part of the slab, a short version is used - with the help of one zigzag line of cup marks, smoothly transitioning from evening to morning ellipses.

The number of cup marks in the southeastern part of the slab coincides with the number of hour marks in this sector (only the 20:30 mark has two cup marks instead of one, possibly due to negligence during application or to mask the measurement technology). The number of cup marks in the southwestern part of the slab significantly exceeds the number of hour markers in this sector. In the range from 3:30 to 6 o'clock in the morning, each hour and half hour mark corresponds to one cup mark. But from midnight to 3:30, instead of seven cup marks, the row consists of 15 cup marks. But, at the same time, directly under the zigzag line of cup marks (on the $X$ axis) there is a row of seven cup marks (Fig. 22).

It seems that it was these cup marks that served to recalculate the time intervals of half an hour from midnight to 3:30 in the morning.

The use of cup marks to recalculate the number of half hours at night could be relevant if time was measured at night using a water clock. And each time the measuring vessel was filled or emptied, a small stone or the like was placed in the corresponding cup mark on the slab.

---

[9] Mathematical (true) horizon.

[10] Calculations of the height of the Sun were made using the program RedShift 7 Advanced.

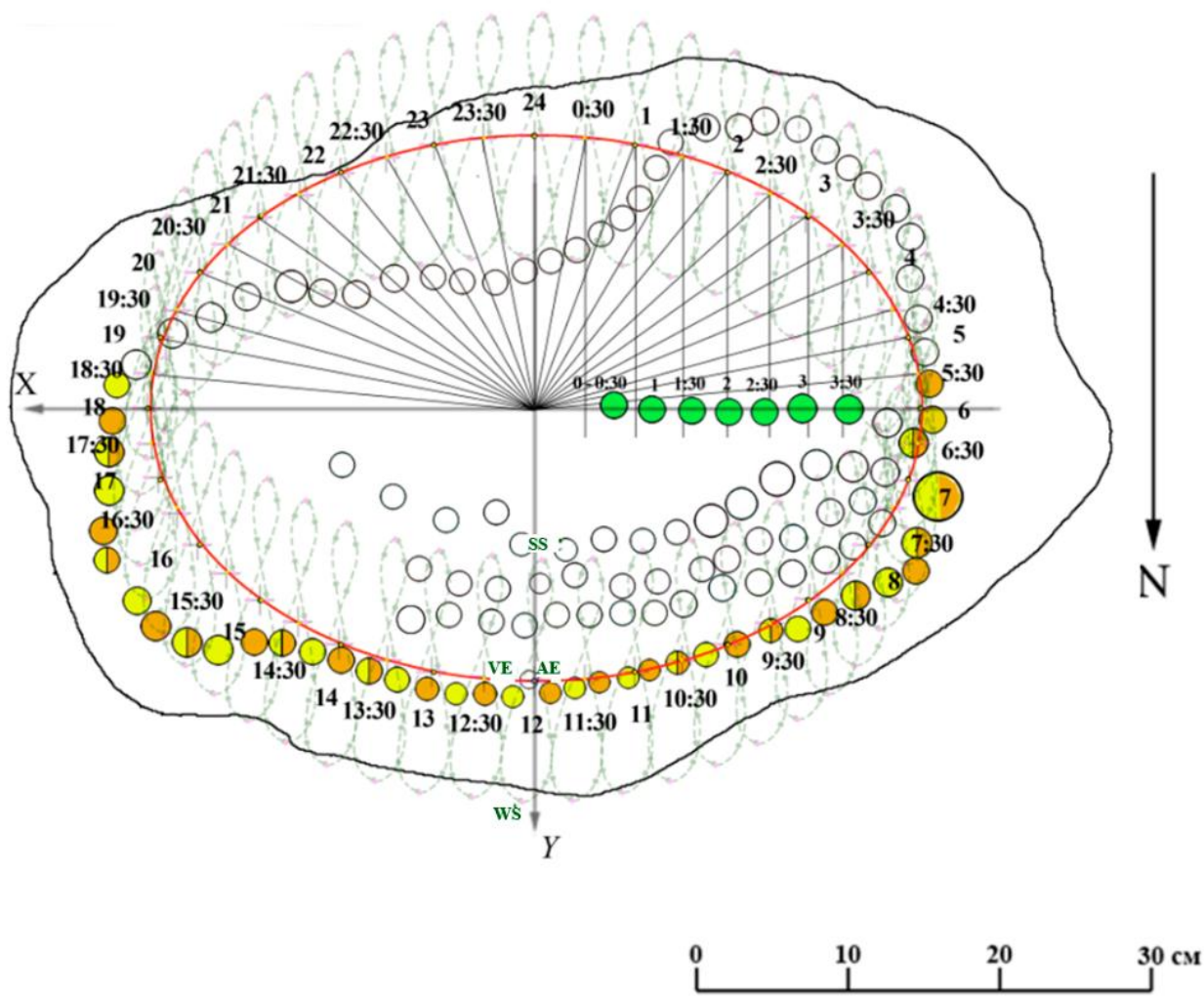

**Figure 22.** Plan-scheme of the Belogorsk slab. Hour lines and dimples of the southern part of the slab, corresponding to the hour markers of the ellipse of the equinoxes. Cup marks for converting the number of half hours from midnight to 3:30 am are highlighted in green.

Thus, as a result of studying the location of cup marks on the surface of the Belogorsk slab, that it is a cadran - a plane with hour divisions of a sundial, which in its type is closest to the analemmatic sundial. However, the principle of hourly marking of the Belogorsk slab is unique and therefore it can be distinguished into a separate type of sundial - inverted analemmatic sundial. This type of sundial is characterized by the fact that, unlike typical analemmatic sundial, the gnomon remains motionless throughout the year, and the "dial" "moves" - an ellipse of hour markers from cup marks, i.e. gnomon and hour markers (cup marks) change places in terms of mobility. The movement, at the same time, is not literal, but is imitated by several rows of cup marks, which are fragments of ellipses of hour markers for different months of the year.

Despite the revolutionary nature of the new design of the sundial and the clear desire to get rid of the movement of the gnomon during the year, thereby simplifying the process of measuring time and increasing its accuracy, the gnomon continued to remain vertical. The next step in the evolution of time measurement technology and the creation of a new type of sundial was reflected on the slab from Popov Yar-2, on which both the markings of traditional analemmatic sundial and qualitatively new horizontal sundial with an inclined gnomon were found (Vodolazhskaya, 2013, p. 68-88). Similar markings were also found on the Staropetrovsky vessel, which also belongs to the Srubnaya culture (Vodolazhskaya, Usachuk, Nevsky, 2015b, p. 43-60). That is, the Belogorsk slab should be dated to an earlier period than the slab from Popov Yar 2, but later than the slab from kurgan field Tavriya, i.e. approximately XIV-XIII centuries BC.

Thus, the idea underlying the sundial on the Belogorsk slab was so revolutionary that we can talk about the discovery of a qualitatively new type of sundial - inverted analemmatic sundial, which existed on the territory of the Northern Black Sea region around the XIV-XIII centuries BC.